\documentclass[useAMS,usenatbib]{mn2e}
\usepackage{float}
\usepackage{graphicx}
\usepackage{grffile}
\usepackage{lineno}
\usepackage{multirow}
\usepackage{subfigure}
\usepackage{color}
\usepackage{cleveref}
\usepackage{deluxetable}
\usepackage{nicefrac}
\usepackage{amsmath,amssymb}
\usepackage{listings}

\def\vFv{\nu F_{\nu}}
\def\Ep{E_{\rm peak}}
\def\Epr{E_{\rm peak}^{\rm rest}}
\def\Epo{E_{\rm peak}^{\rm obs}}
\def\FE{F_{\rm E}}
\def\dL{d_{\rm L}(z)}
\def\dLsq{d_{\rm L}^2 (z)}
\def\Nr{N_{\rm rest}}

\def\mean#1{\left< #1 \right>}

\title[The Rest-Frame Golenetskii Correlation]{The Rest-Frame Golenetskii Correlation via a Hierarchical Bayesian Analysis }\author[J. Michael
Burgess]{J. Michael Burgess$^{1,2}$\thanks{E-mail: jamesb@kth.se (JMB)}\\
  $^{1}$The Oskar Klein Centre for Cosmoparticle Physics,
  SE-106 91 Stockholm, Sweden\\
  $^{2}$Department of Physics, KTH Royal Institute of Technology,
  AlbaNova, SE-106 91 Stockholm, Sweden}
\begin{document}

\date{Accepted XXXX December XX. Received XXXX December XX; in original form XXXX October XX}

\pagerange{\pageref{firstpage}--\pageref{lastpage}} \pubyear{2014}

\maketitle

\label{firstpage}

\begin{abstract}

  Gamma-ray bursts (GRBs) are characterised by a strong correlation
  between the instantaneous luminosity and the spectral peak energy
  within a burst. This correlation, which is known as the
  hardness-intensity correlation or the Golenetskii correlation, not
  only holds important clues to the physics of GRBs but is thought to
  have the potential to determine redshifts of bursts. In this paper,
  I use a hierarchical Bayesian model to study the universality of
  the rest-frame Golenetskii correlation and in particular I assess
  its use as a redshift estimator for GRBs.  I find that, using a
  power-law prescription of the correlation, the power-law indices
  cluster near a common value, but have a broader variance than
  previously reported ($\sim 1-2$). Furthermore, I find evidence that
  there is spread in intrinsic rest-frame correlation normalizations
  for the GRBs in our sample ($\sim 10^{51}-10^{53}$ erg
  s$^{-1}$). This points towards variable physical settings of the
  emission (magnetic field strength, number of emitting electrons,
  photospheric radius, viewing angle, etc.).  Subsequently, these
  results eliminate the Golenetskii correlation as a useful tool for
  redshift determination and hence a cosmological probe. Nevertheless,
  the Bayesian method introduced in this paper allows for a better
  determination of the rest frame properties of the correlation, which
  in turn allows for more stringent limitations for physical models of
  the emission to be set.

\end{abstract}


\section[]{Introduction}
A key feature of gamma-ray bursts (GRBs) is the observed relation
within a burst between the luminosity and the $\vFv$ peak energy
($\Ep^{\rm rest}$)
\citep{Golenetskii:1983,Kargatis:1994,Borgonovo:2001,Ryde:2006,Ghirlanda:2010,Lu:2010bj,Zhang:2012fq,Guiriec:2013hl,Burgess:2014}. This
luminosity-$\Ep^{\rm rest}$ relation was first discovered by
\citet{Golenetskii:1983} and has been found in several GRBs regardless
of their lightcurve shape. The relation is sometimes referred to as
the hardness-intensity correlation or the Golenetskii correlation
(GC). It is typically stronger during the decay phase of a GRB
lightcurve. The form of the GC states that the luminosity of the GRB
is proportional to its $\Ep^{\rm rest}$ to some power $\gamma$:

\begin{equation}
  \label{eq:ler}
  L \propto \left( \Ep^{\rm rest} \right)^{\gamma} \; {\rm erg\; s}^{-1}
\end{equation}
 
\noindent Historically, the GC has been used in an attempt to
understand the physical process generating the observed emission
\citep[e.g.][]{Dermer:2004,Ryde:2006,Bosnjak:2014}. \citet{Ryde:2006}
pointed out that the correlation is strongest in GRBs with single
non-overlapping pulses.

Such a correlation should be a signature of the evolving radiative
process occurring in an outflow. For example, if the emission is
purely photospheric then one would expect $L\propto\sigma_{\rm B}
T^{4}$ where $\sigma_{\rm B}$ is the Boltzmann constant and $\Ep
\propto 3T$ where $T$ is the temperature of the outflow plasma. This
fact motivated the original research into photospheric emission of
GRBs. \citet{Borgonovo:2001,Ryde:2006} used single- and
multi-component spectral models composed of thermal and non-thermal
components to analyze several GRBs and obtain their GCs. It was found
that the GCs of these bursts exhibit a wide range of $\gamma$
\citep[see also][]{Burgess:2014} and it is difficult to assign a
common physical setting to all GRBs: it is known that photospheric
emission can take on several forms due to the differences in viewing
angle \citep{Lundman:2014} as well as subphotospheric dissipation
\citep[e.g.][]{Peer:2005,Beloborodov:2010} or jet composition
\citep{Giannios:2006,Begue:2015}. These advanced photospheric models
have just begun making predictions to explain the GC
\citep{Fan:2012,Lopez-Camara:2014}.

Other attempts to explain the observed GCs invoked non-thermal
synchrotron emission \citep{Zhang:2002dr,Dermer:2004} in both single
\citep{Ghirlanda:2010} and multi-component
\citep{Burgess:2014,Preece:2014} time-resolved spectra. For
synchrotron emission, one expects $L \propto N_{\rm e} B^2 \Gamma^2
\gamma_{\rm e}^2$ and $\Ep \propto B \Gamma \gamma_{\rm e}^2$ where
$N_{\rm e}$ is the number of emitting electrons, $B$ in the magnetic
field, $\gamma_{\rm e}$ is the characteristic electron Lorentz factor
and $\Gamma$ is the bulk Lorentz factor. This implies that $L \propto
N_{\rm e} \Ep^2 \gamma_{\rm e}^{-2}$. Deeper considerations for the evolution of the
parameters can lead to differenent $\gamma$'s (see \citet{Dermer:2004}
for examples.) Yet again, the wide range of observed $\gamma$'s make
it difficult to explain all GRBs with synchrotron emission.

Another feature of the GC exploits the fact that if the relation is
generated by a common process, then GCs can be used to estimate
redshift of GRBs \citep{Guiriec:2013hl,Guiriec:2015}. Several GRBs
with known redshift exhibited a common rest-frame proportionality
constant ($N\simeq 10^{53}$ erg s$^{-1}$) between luminosity and
$\Ep^{\rm rest}$. If this is true for all GRBs, it could be possible
to use the observer-frame normalization of the relation to estimate
the redshifts of GRBs that do not have one measured by other
means. This estimation of redshift differs from that of the so-called
Amati and Yonetoku relations \citep{Amati:2002,Yonetoku:2004}. These
relations rely on the time-integrated properties of GRBs and most
likely result from the effects of functional correlation
\citep{Massaro:2007} and selection effects \citep{Kocevski:2012}. If
time-resolved GCs such as those studied herein do allow for the
estimation of redshift then this would allow for a very powerful
cosmological tool as GRBs can probe the high-redshift
universe. However, this possibility heavily relies on the assumption
of a common rest-frame normalization, which in both the photospheric
and synchrotron models is not predicted.

\section[]{The Relation and the Bayesian Model}

\label{sec:model}

The GC is given in the rest-frame in Equation \ref{eq:ler}; however,
it is derived from observer-frame time-resolved spectral fits of
GRBs. The observed spectra are integrated over energy (10 keV - 40 MeV) to yield the
energy flux ($\FE$) which, assuming isotropic emission, is related to
the luminosity via $L = 4 \pi \dLsq \FE$.  Here, $z$ in the redshift
of the GRB and $\dL$ is the luminosity distance. Similarly, $\Ep^{\rm
  rest} =\Ep^{\rm obs} (1+z)$.

Assuming that the time-evolving rest-frame luminosity of GRBs derives
from a single inherent relationship given as

\begin{equation}
  \label{eq:1}
  L=\Nr \left(\frac{\Ep^{\rm rest}}{100 {\rm keV}} \right)^{\gamma} \; {\rm erg\; s}^{-1}{\rm ,}
\end{equation}
\noindent where $\Nr$ is a normalization attributed to the intrinsic
physics of the emission process, and $\gamma$ is the GC index again
attributed to an intrinsic physical process, we shift into the
observer-frame such that

\begin{equation}
  \label{eq:3}  
\FE = \frac{\Nr}{4 \pi \dLsq} \left( \frac{\Ep^{\rm obs} (1+z)}{100 {\rm keV}}  \right)^{\gamma} \; {\rm erg\; s}^{-1} {\rm \; cm}^{-2}{\rm .}
\end{equation}

\noindent Taking the base-10 logarithm to obtain a linear relationship we have

\begin{equation}
  \label{eq:5}
  \log(\FE) = \log(\Nr) - \log(4 \pi \dLsq) + \gamma  \log\left(\frac{\Ep^{\rm rest}(1+z)}{100 {\rm keV}}\right) {\rm .}
\end{equation}

\noindent If one makes the assumption that $\Nr$ and $\gamma$ remain
constant or tightly distributed for all GRBs, from this equation one
can solve for $z$. I use this assumption to construct a hierarchical
Bayesian model \citep[see for example, ][for related uses of
hierarchical Bayesian models in
astrophysics]{Mandel:2011,March:2011jx,Andreon:2012dp,Sanders:2015,deSouza:2015gk}. For
the $i^{\rm th}$ GRB$_i$ we have

\begin{equation}
  \label{eq:lin}
  \log\left(\FE^{i,j}\right) = \log\left( \Nr \right) - \log \left(4 \pi d_{\rm L}^2 \left( z^i \right) \right) + \gamma^{i} \log\left(\frac{\Ep^{{\rm obs}i,j}  \left( 1+z^i \right)}{100 {\rm  keV}}\right) {\rm .}
\end{equation}

\begin{figure}
  \centering
  \includegraphics[scale=1]{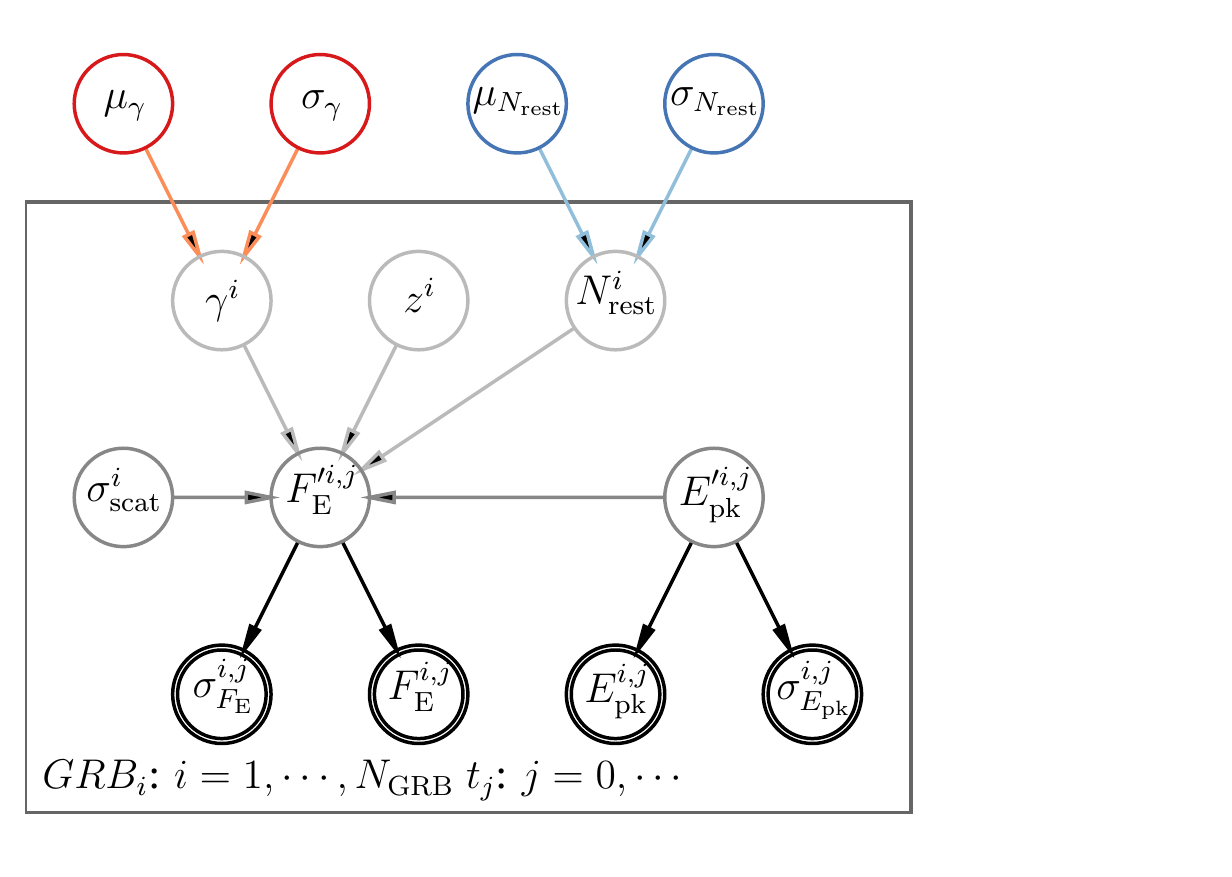}
  \caption{Graphical representation of the Bayesian hierarchical
    models. For visual simplicity, logarithmic quantities in the
    linear Equation \ref{eq:lin} are represented as $Q=\log(Q)$ as
    well as with their associated priors. Here, $i$ runs over the
    number of GRBs ($N_{\rm GRB}$) in the data set and $j$ is the
    time-evolution iterator. \textit{Mod~A} is the full model while
    \textit{Mod~B} and \textit{Mod~C} are represented by disregarding
    the \textit{red} and then the \textit{blue} hyper-priors
    respectively.}
  \label{fig:model}
\end{figure}

\noindent Here, $j$ indexes time. Figure \ref{fig:model} demonstrates
the model. I will call this model \textit{Mod~A}. Henceforth,
logarithmic quantities ($\log\left( \Nr \right)$, $\log\left( \FE
\right)$, etc.)  will be represented as $Q=\log\left( Q \right)$
for notational simplicity. Since I choose $\FE$ as the dependent
quantity and $\Epo$ as the independent quantity, the measurement error
in $\Epo$ must be accounted for following the methods of
\citet{Dellaportas:1995,Andreon:2013}. I also allow for intrinsic
scatter in each GC. The following priors are assumed:

\begin{eqnarray}
\mu_{\gamma} & \sim & \mathcal{N}(0,\text{std}(\FE)/\text{std}(\Epo)\\
\sigma_{\gamma} & \sim & \textit{Cauchy}_{(0,\infty)}(0,2.5)\\
\gamma^i &\sim&  \mathcal{N}(\mu_{\gamma},\sigma_{\gamma})\\
\mu_{\Nr} & \sim & \mathcal{N}(52,5)\\
\sigma_{\Nr} & \sim & \textit{Cauchy}_{(0,\infty)}(0,2.5)\\
\Nr^i &\sim&  \mathcal{N}(\mu_{\Nr},\sigma_{\Nr})\\
z^i &\sim& \mathcal{U}(0,15)\\
\Ep^{{\rm obs} \prime i,j} &\sim& \mathcal{N}(\Ep^{{\rm obs} i,j},\sigma_{\Ep}^{i,j})\\
\sigma_{\rm scat}^{2,i} &\sim& \textit{Cauchy}_{(0,\infty)}(0,2.5)\\
\FE^{\prime i,j} &\sim& \mathcal{N}(\star,\sigma_{\rm scat}^i)\\
\FE^{ i,j} &\sim& \mathcal{N}(\FE^{\prime i,j},\sigma_{\FE}^{i,j})\\
\end{eqnarray}

\noindent Here ($\star$) indicates the right-hand side of Equation
\ref{eq:lin}. Half-Cauchy distributions were chosen as non-informative
proper priors for the variances \citep{Gelman:2006di}. If a GRB has a
known redshift, then it can be fixed in the model which will aid in
constraining the hyper-parameters for $\Nr$.

I can further generalize the model by relaxing the assumption of a
common $\Nr$ and $\gamma$ and allow separate $\Nr$ and $\gamma$ to
be fit without their respective hyper-priors. Therefore, I implement
two further models, \textit{Mod B} and \textit{Mod~C} which relax the
assumption of common $\gamma$ and common $\Nr$ respectively. These
models are also detailed in Figure \ref{fig:model} when the
\textit{red} and then the \textit{blue} hyper-priors are removed.

The models are implemented in {\tt
  PyStan} \citep{stan} which is a probabilistic
modeling language implementing a Hamiltonian Monte Carlo sampler for
full Bayesian inference. The posterior marginal distributions of the
parameters obtained from {\tt PyStan} can be analyzed to determine
$\gamma^i$, $\Nr^i$, and $z^i$. The simultaneous fitting of all GCs
with linked parameters (e.g. $\gamma^i$) of the hierarchical model
provides shrinkage of the estimates and, if the data support it, will
pull together the estimates as well as assist in better fits of GCs
where there is less data.  For computational speed, I implement the
highly accurate, analytic form of $\dL$ from \citet{Adachi:2012bn} in
my model.

An alternative approach to determining redshifts is to fit all GCs
from GRBs with known redshift together as one data set via a standard
fitting tool such as {\tt FITEXY} \citep{Press:2007}. A common,
calibration rest-frame normalization ($N_{\rm cal}$) and slope can be
determined. Then, GRBs without known redshift can have their
observer-frame GCs fitted to determine their observer-frame
normalizations ($M_z$) which can be solved for redshift (see Section
\ref{sec:simA}). However, this approach neglects the robustness of the
Bayesian model presented here because in the Bayesian model, all
parameters are determined from the data simultaneously accounting for
all correlations in the parameters and data.

\section{Simulations}

To test the feasibility of this approach, I simulate a set of eight
GCs to fit with the model. Starting with a universal relation

\begin{equation}
  \label{eq:2}
  L^{\rm sim} = \Nr^{\rm sim} \left( \frac{\Ep^{\rm rest, sim}}{100 {\rm keV}}  \right)^{\gamma^{\rm sim}}  \; {\rm erg\; s}^{-1}
\end{equation}

\noindent a random number of $\Ep^{\rm rest, sim}$'s are drawn
from a log-uniform distribution. For \textit{Mod~A}, $\Nr^{\rm
  sim}=52$ erg s$^{-1}$ and $\gamma^{\rm sim} =1.5$ for every
synthetic GRB (See Figure \ref{fig:simAdata}). For \textit{Mod~B} and
\textit{Mod~C}, $\gamma^{\rm sim}$ is drawn from $\mathcal{U}(1,2)$ and for
\textit{Mod~C} $\Nr$ is drawn from $\mathcal{U}(51.7,52.3)$ (see
Figures \ref{fig:simBdata} and \ref{fig:simCdata}). Then, $L^{\rm
  sim}$ is computed for each $\Ep^{\rm rest, sim}$. A random redshift
is assigned and the rest-frame quantities are shifted into the
observer-frame. Finally, each observer frame quantity has
heteroscedastic error added to it assuming that the noise is normally
distributed for both $\Ep^{\rm obs, sim}$ and $\FE^{\rm sim}$. No
intrinsic scatter is added.

\subsection{Simulated GCs with Common $\Nr$ and $\gamma$: \textit{Mod~A}}

\label{sec:simA}

The set of simulated GCs is fit with \textit{Mod~A} assuming that
$GRB_1$ - $GRB_4$ have known redshifts and the remaining GRBs do
not. Table \ref{tab:simA} details the fits and marginal posterior
distributions for all parameters are displayed in Appendix
\ref{sec:post}. The observer-frame relations are well recovered by the
model and can be seen in Figure \ref{fig:simAfits}. The estimated
$\gamma^i$  are well recovered by the model. The $\Nr^{i}$ are
also well fitted even for GRBs that do not have a redshift owing to
the shrinkage of the posterior by GRBs with known redshift. There is a
tight correlation between $\mu_{\gamma}$ and $\mu_{\Nr}$ (Figure
\ref{fig:simAmu}) for the hyper parameters. The estimated unknown
redshifts have long tails but are generally estimated well as shown in
Figure \ref{fig:simzcomp}. All of the simulated values are within the
95\% highest posterior density intervals (HDIs). Still, the spread in
estimated values makes them unusable as a cosmological probe.

Now, I compare to the frequentist approach to see the difference in
the methods. This approach requires at least one redshift be known to
calibrate the $\Nr$. The same four simulated GCs are used as known
redshift GRBs as in the previous sections. The method proceeds in the
following fashion:
\begin{itemize}
\item The four known redshift GRBs have their \textit{rest-frame} GCs
  fit via the {\tt FITEXY} method to determine a calibration value for
  $\Nr$.

\item The GRBs without redshift have their \textit{observer-frame}
  GCs fit to determine their normalization which is
  $A=\nicefrac{M_z}{\Nr}$ where $M_z$ is the observer-frame
  normalization.

\item Using the calibration $\Nr$, $M_z$ is solved for and
  subsequently $z$ is obtained.
\end{itemize}

Table \ref{tab:mleA} details the results from the procedure. The
calibrated rest-frame normalization is $N_{\rm cal}=52 \pm 0.01$ with slope
$\gamma=1.49 \pm 0.014$ (compare these to the simulated values of
$\Nr=52$ and $\gamma = 1.5$) It is important to note that {\tt FITEXY}
has a positive bias on the obtained $\gamma$'s, an effect that is well
known \citep[e.g.][]{Kelly:2007} and the method's statistical
properties are poorly understood. This makes the fitting GCs via the
method unreliable at determining the inherent physics in the outflow.

The 1$\sigma$ errors on redshift are determined numerically by fully
propagating the errors from the linear fits of both the calibration
and observer-frame fits. The results are displayed in Table
\ref{tab:mleA} The hierarchical Bayesian model performs better at
obtaining the simulated $\gamma$'s (see Table \ref{tab:simA}). The
redshifts are obtained accurately via {\tt FITEXY} but with very small
errors that do not take into account the full variance of the model
and data. Therefore, these errors are likely underestimated. We will
see how these under-estimated errors lead to inaccurate redshift
predictions in the following sections.

\subsection{Simulated GCs with Varying $\gamma$: \textit{Mod~B}}

For simulations where $\gamma^i$ is varied, \textit{Mod~B} is used to
fit the data. Similar to \textit{Mod~A}, \textit{Mod~B} is able to
estimate the simulated properties of the GRBs. Table
\ref{tab:simB} details the results. All parameter marginal posterior
distributions are shown in Appendix \ref{sec:post}. Importantly, the
varying values of $\gamma^{i}$ are well estimated and the redshifts
are found but with higher HDIs than those found with \textit{Mod~A}
(See Figure \ref{fig:simzcomp}).

I then proceed to fit the simulated data with {\tt FITEXY}. The
results are in Table \ref{tab:mleB} and show that the redshifts are
inaccurately estimated and the errors are underestimated. Therefore,
if there is a distribution of $\gamma^i$ in real data, this method
will give incorrect redshifts but the errors may not encompass the
true values..

\subsection{Simulated GCs with Varying $\gamma$ and $\Nr$: \textit{Mod~C}}

Finally, I test \textit{Mod~C} which has no linkage between
datasets. The fits are detailed in Table \ref{tab:simC} and all
parameter marginal posterior distributions are displayed in Appendix
\ref{sec:post}. GRBs with known redshift can easily have $\Nr$
estimated but the degeneracy between $\Nr$ and $z$ for GRBs without
known redshift restricts the ability for $\Nr$ to have a compact
marginal distribution. As a consequence, Figure \ref{fig:simCz} shows
that redshift cannot be determined in this model. The model is still
able to determine $\gamma^i$, though, with broader HDIs. The loss of
linkage and information sharing between GRBs with known and unknown
redshift in this model makes it not useful for cosmology though if it
is the model representing the actual physics of GCs, it can be used on
GRBs with known redshift to determine rest-frame properties. Once
again, {\tt FITEXY} is tested (Table \ref{tab:mleC}) and it is found
that the redshifts are inaccurately estimated with underestimated
errors.

\subsection{Reconstructing the Simulated Golenetskii Correlation}
\label{sec:simgol}

Another test of the models is how well they reconstruct the simulated
relations in the rest-frame. Essentially, the marginalization over
redshift removes the cosmological factors in the GCs and allows for
the rest-frame quantities to be calculated with the full variance of
the model and data. This is more descriptive than simply shifting the
observer-frame quantities into the rest-frame using the estimated and
known redshifts.

The process of calculating the rest-frame Golenetskii relation depends
on whether or not the redshift for a GRB is known. For GRBs with
unknown redshift, I first construct the observed rest-frame
$\Epr$'s. Since I model the measurement error of the observer-frame
$\Epo$ during the fits, I obtain a distribution for each $\Epo$ based
off the data and the model. I propagate the distribution of $\Epo$
and the estimated redshift ($z_{\rm est}$) to reconstruct $\Epr = \Epo(1+z_{\rm
  est})$. Thus a distribution for $\Epr$ is obtained. Similarly, I
propagate the obtained distributions of the marginalized parameters to
reconstruct $L = \Nr \left( \nicefrac{\Epr}{100 {\rm
      keV}}\right)^{\gamma}$.

\begin{figure*}
  \centering

\subfigure[]{\includegraphics{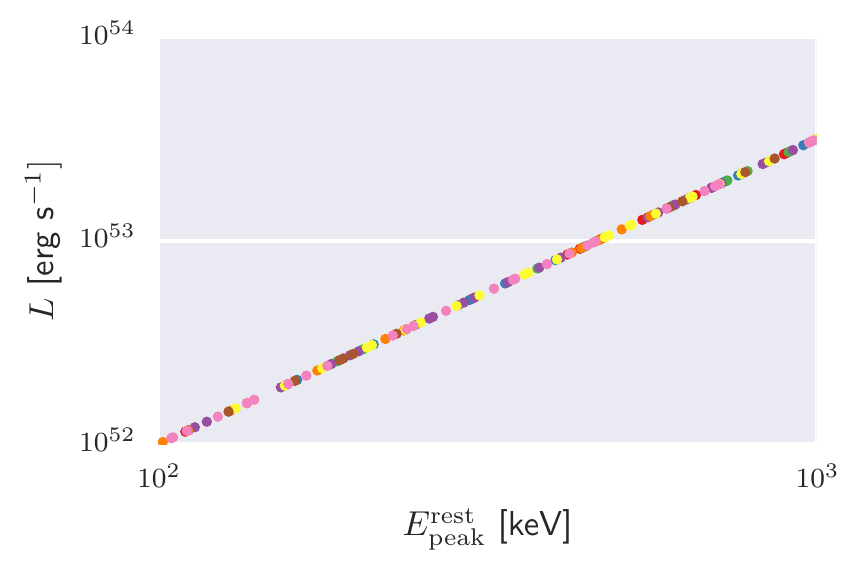}} \subfigure[]{\includegraphics{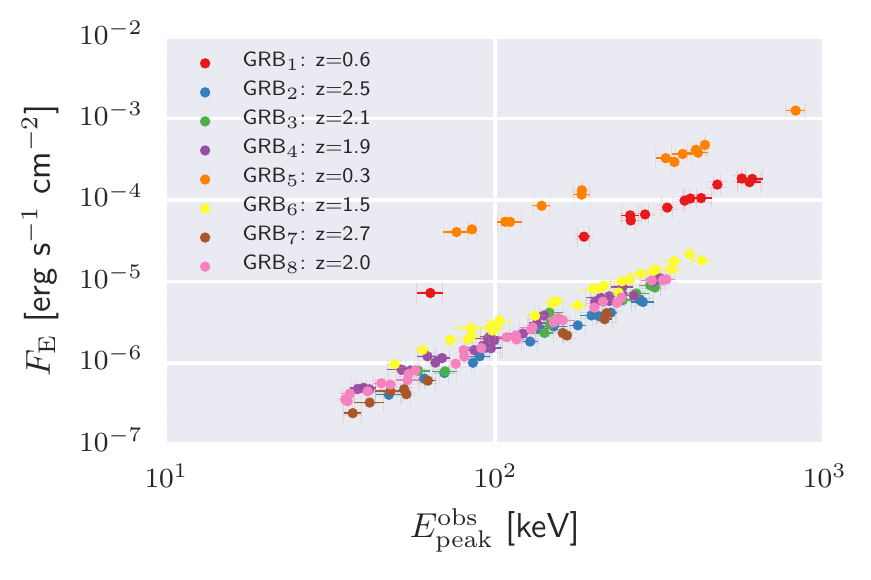}}

\caption{The simulated rest-frame (left) and observer-frame (right)
  data from \textit{Mod~A}}
  \label{fig:simAdata}
\end{figure*}

\begin{figure*}
  \centering

\subfigure[]{\includegraphics{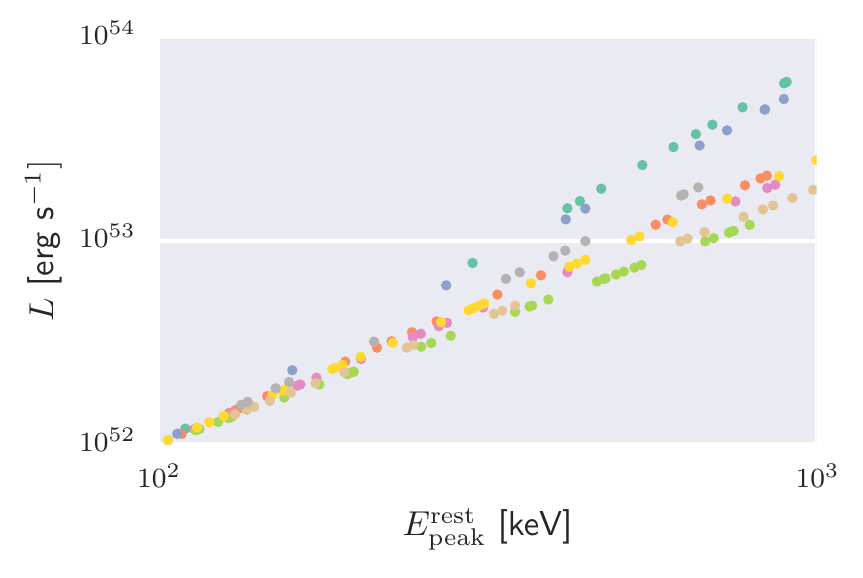}} \subfigure[]{\includegraphics{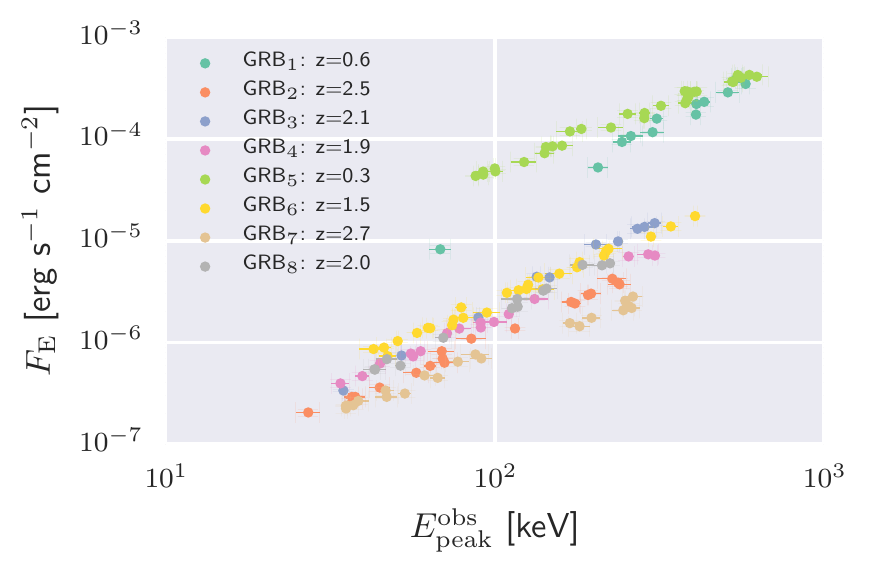}}

\caption{The simulated rest-frame (left) and observer-frame (right)
  data from \textit{Mod~B}}
  \label{fig:simBdata}
\end{figure*}

\begin{figure*}
  \centering

\subfigure[]{\includegraphics{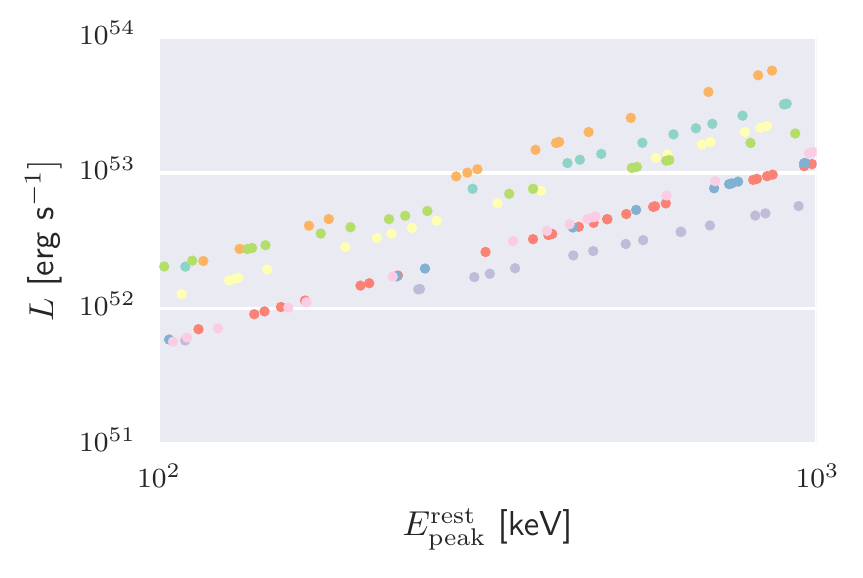}} \subfigure[]{\includegraphics{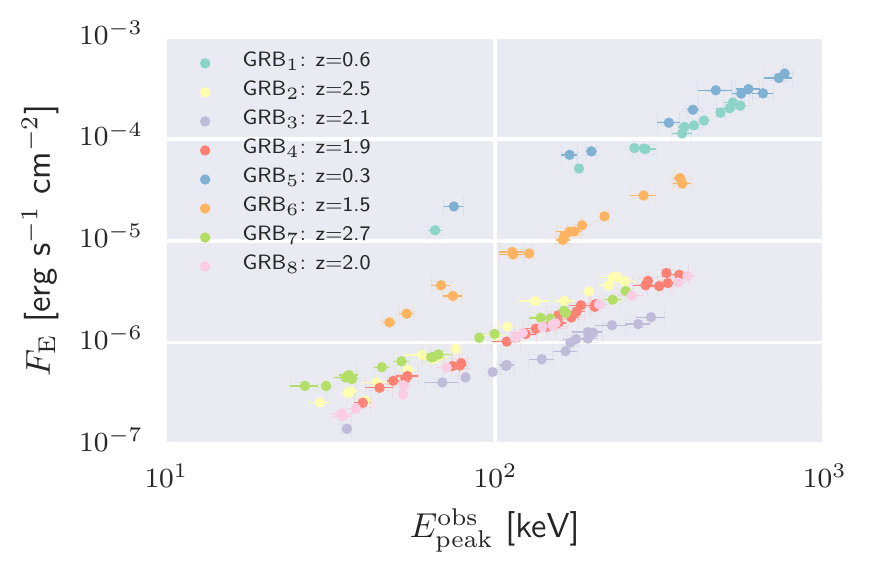}}

\caption{The simulated rest-frame (left) and observer-frame (right)
  data from \textit{Mod~C}}
  \label{fig:simCdata}
\end{figure*}

\begin{figure}
  \centering
  \includegraphics{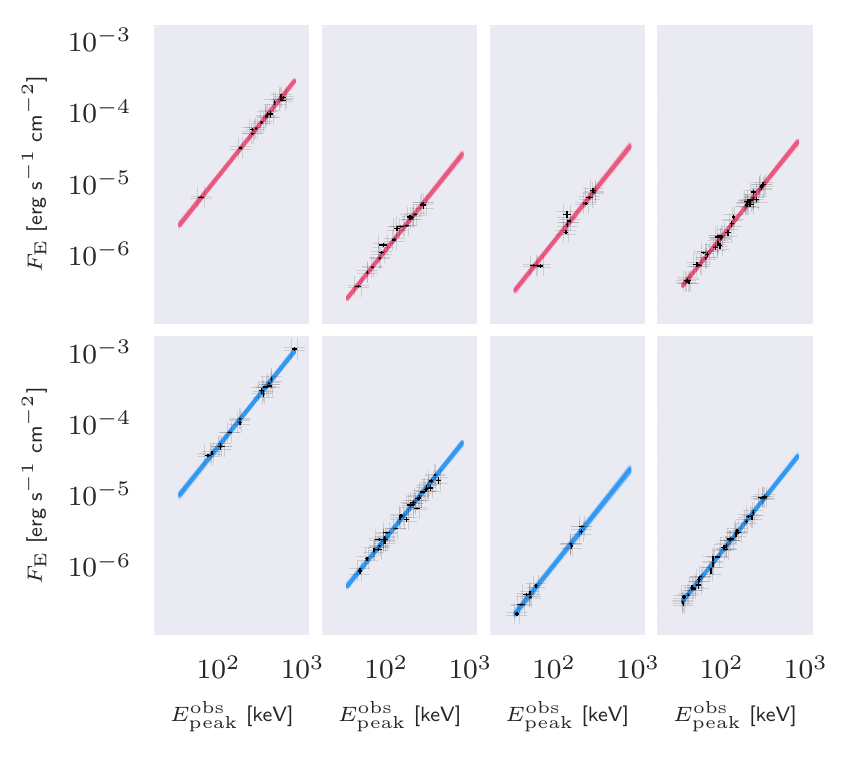}
  \caption{The reconstructed observer-frame GCs from simulations of
    \textit{Mod~A} assuming four redshifts are known. The \textit{red}
    lines map out the posterior of known redshift GRBs while
    \textit{blue} maps out the posterior of unknown redshift GRBs. The
    plots correspond to GRB$_1$ - GRB$_8$ from top-left to bottom
    right respectively.}
  \label{fig:simAfits}
\end{figure}

\begin{figure}
  \centering
  \includegraphics{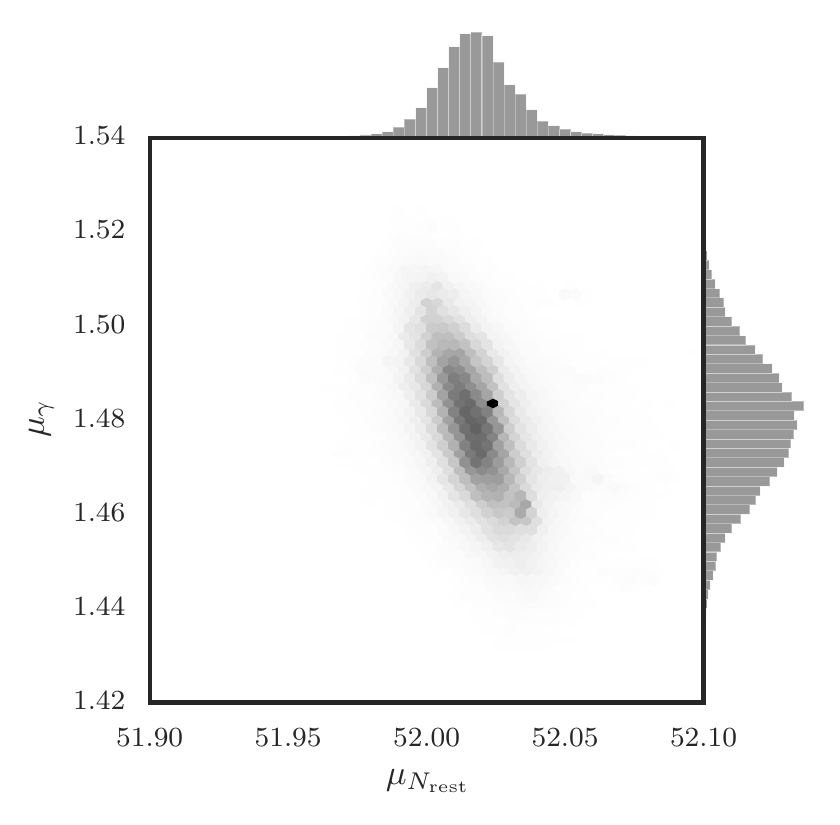}
  \caption{The distributions of hyper-parameters for $\Nr^i$ and
    $\gamma^i$ for simulations of \textit{Mod~A}.}
  \label{fig:simAmu}
\end{figure}

\begin{figure}
  \centering
  \includegraphics{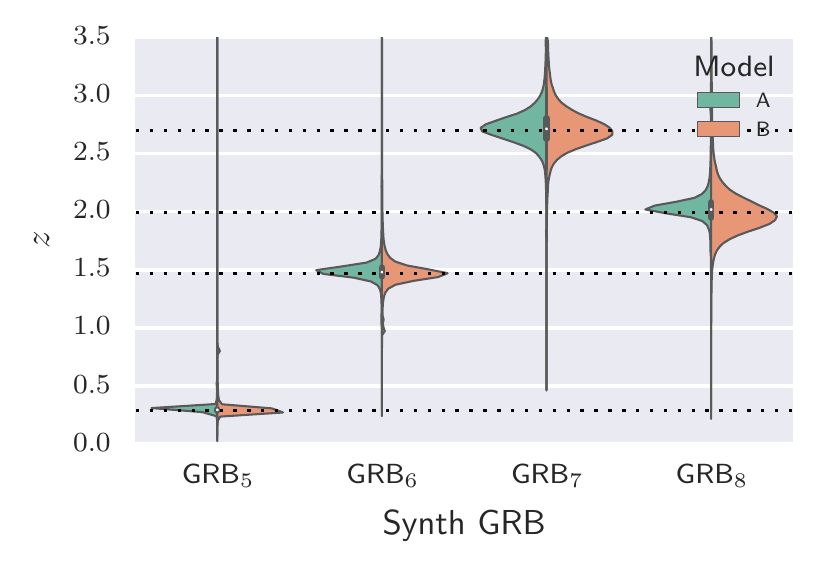}
  \caption{Comparing the redshift estimations of \textit{Mod~A} and
    \textit{Mod~B}. In general, \textit{Mod~A} has tighter constraints
    on redshift estimation. The dashed lines indicate the simulated
    values.}
  \label{fig:simzcomp}
\end{figure}

\begin{figure}
  \centering
  \includegraphics{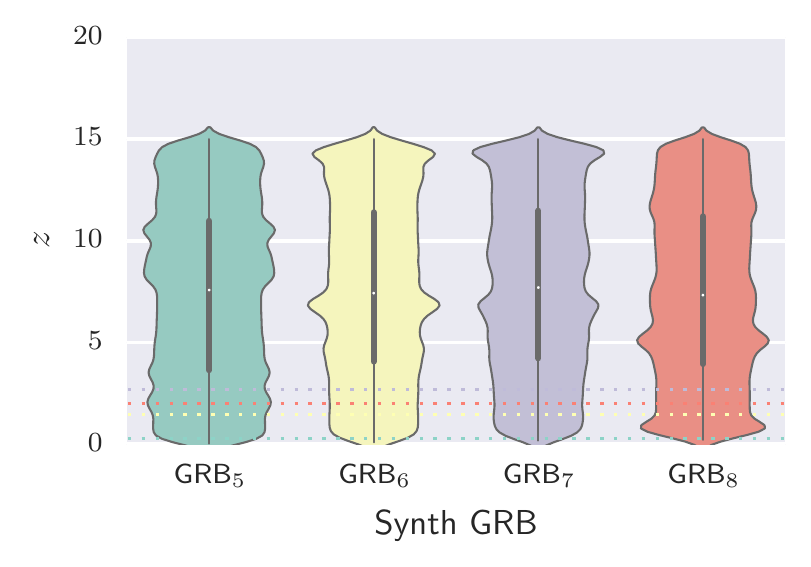}
  \caption{The estimated distributions of unknown redshifts for
    simulations of \textit{Mod~C}. The dashed lines indicate the
    simulated values.}
  \label{fig:simCz}
\end{figure}

\begin{figure}
  \centering

\subfigure[]{\includegraphics{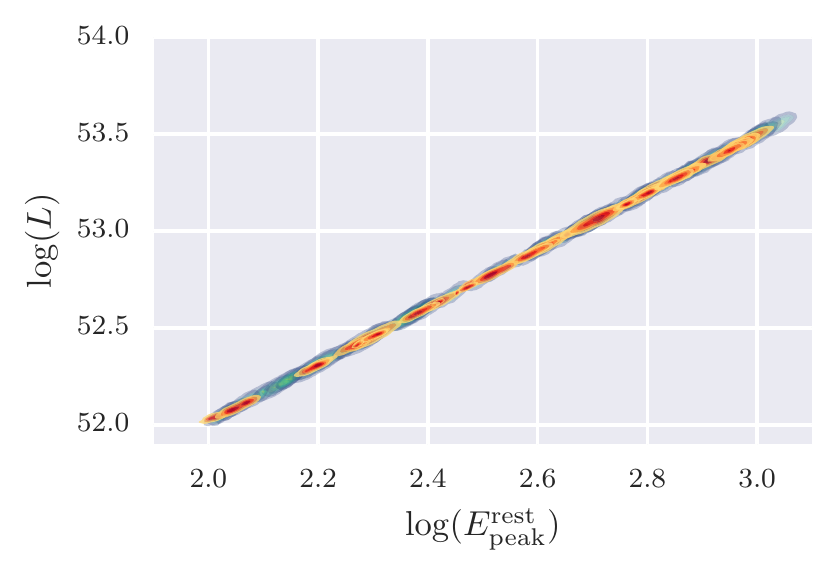}}
\subfigure[]{\includegraphics{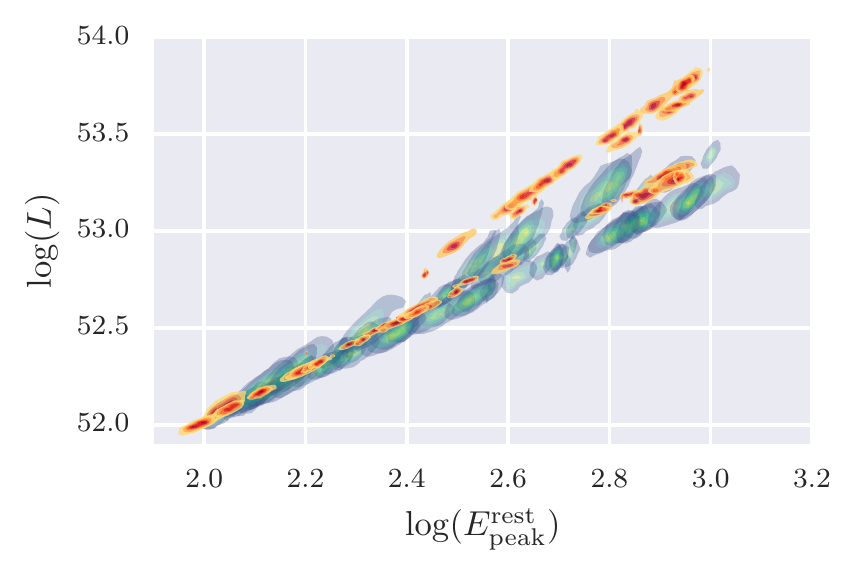}}
\subfigure[]{\includegraphics{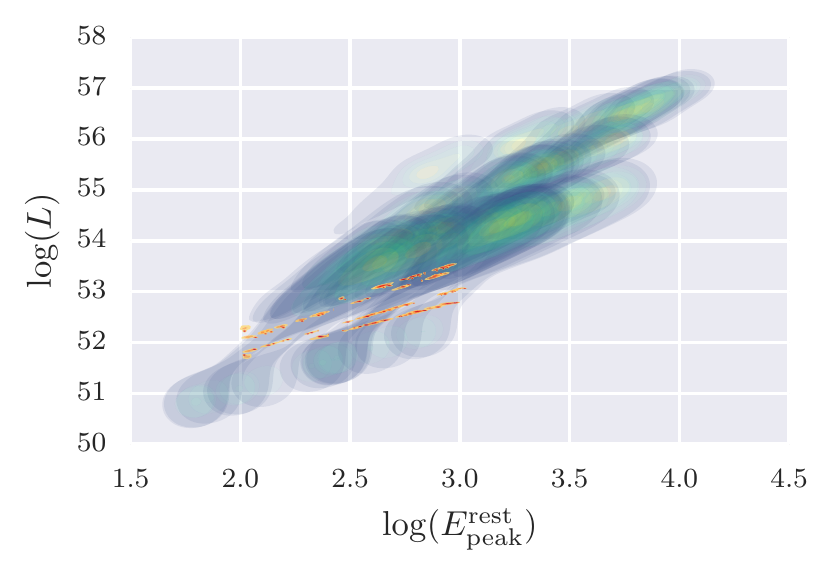}}

\caption{The reconstructed rest-frame Golenetskii correlations from
  data simulated and fit with (from top to bottom) \textit{Mod~A},
  \textit{Mod~B}, and \textit{Mod~C}. The fits reconstruct the
  simulated relations (Figures \ref{fig:simAdata} - \ref{fig:simCdata})
  for both GRBs with known (\textit{yellow-red}) and unknown
  (\textit{blue-green}) redshift excellently with the exception of
  \textit{Mod~C} which cannot determine the relation for GRBs without
  known redshift.}
  \label{fig:simGol}
\end{figure}

For GRBs with known redshift, the process is the same except the
measured value of $z$ is used rather than the estimated value. At the
end of the process, I have distributions of $L$ and $\Epr$ for each
data point of each GRB. Figure \ref{fig:simGol} displays the results
of this process for all three models. Compared with the simulated
relations (Figures \ref{fig:simAdata} - \ref{fig:simCdata}) all models
reconstruct the relation well for both GRBs with known and unknown
redshift except \textit{Mod~C} which cannot estimate unknown
redshifts.

\section{Application to Real GRBs}
\label{sec:real}

The model is now applied to real data from a sample GRBs that have
been analyzed with a multi-component spectral model. It has been
claimed that the use of multi-component spectra tighten the observed
GC in some GRBs \citep{Guiriec:2015}. I therefore use the
time-resolved luminosity and $\Ep$ from Band function
\citep{Band:1993} fits of the sample of GRBs analyzed in
\citet{Burgess:2014}. These GRBs are single pulsed in their
lightcurves and were analyzed with physical synchrotron models as well
as the empirical Band function. Additionally, the models were analyzed
under a multi-component model consisting of a non-thermal component
modeled as a Band function and a blackbody. I choose from this sample
GRBs 081224A, 090719A, 090809A, 110721A, and 110920A. Three tentative
redshifts for GRB 110721A appear in the literature ($z=0.382, 3.2,
3.512$) \citep{Berger:2011,Griener:2011}, though none are stringent
measurements. Nonetheless, I will assume that the actual value is
$z=3.2$ for consistency with \citet{Guiriec:2015}.

In \citet{Guiriec:2015}, GRBs were also analyzed with a
multi-component model including GRB 080916C which has measured
redshift of $z=4.24$. I reanalyzed this GRB with the model and
time-intervals posed in \citet{Guiriec:2015} and added this to my
sample. The nearby ($z=0.34$) GRB 130427A is included as well because
of its brightness which allows for several time-intervals to be fit
with spectral models. Though in \citet{Preece:2014} the GRB was fit
with both synchrotron and Band models, only the fits with synchrotron
statistically required a blackbody. Therefore, fits from the Band
function only are used in this work. Finally, I include
multi-component analysis of GRB 141028A which has a well measured
redshift of $z=2.332$ \citet{Burgess:2015}.

All spectral analysis was carried out with {\tt
  RMFIT}\footnote{http://fermi.gsfc.nasa.gov/ssc/data/analysis/rmfit/}. The
$\FE$ was calculated for the Band function for each time-interval
using full error propagation of all spectral parameters including
those of the other spectral components to obtain the errors on the
$\FE$ (See Appendix \ref{sec:prop} for a discussion on error
propagation). I use only Band function fits from the multi-component
fits because it is claimed that they provide a better redshift
estimator. Finally, it is known that GCs are stronger and possess
positive slope in the decay phase of GRB lightcurves. They are
typically anti-correlated in the rising portion of the lightcurve,
therefore, the rise-phase GC is disregarded for this analysis. With
this sample (see Figure \ref{fig:realdata}), I proceed with testing
the Bayesian models.

\subsection{Results}

First, I fit the real GCs with \textit{Mod~A}. Table \ref{tab:realA}
and Figure \ref{fig:realAfits} show the results of the fits. Figure
\ref{fig:realAgamma} shows some variation in $\gamma^i$, but the
values are clustered due to the pull of the hyper-parameters of the
model. Most importantly, GRBs without known redshift have
unconstrained $\Nr^i$ resulting in unconstrained redshifts (see
Figures \ref{fig:realANr} and \ref{fig:realAz}).The hyper-parameters
from the fits are clustered (Figure \ref{fig:realAmu}) due mainly to
GRBs with known redshift.

Next, \textit{Mod~B} is fit to the data. Table \ref{tab:realB} and
Figure \ref{fig:realBfits} show the results of the fits. The values of
$\gamma^i$ vary due to the loosening of their hyper-parameter
constraint (see Figure \ref{fig:realBgamma}). Additionally, estimates of $\Nr^i$
have tighter constraints than with \textit{Mod~A}, yet they are still
broad (See Figure \ref{fig:realBNr}). However, redshift estimation is
still unconstrained though the distributions are peaked at lower
redshifts with heavy tails (see figure \ref{fig:realBz}).

Finally, \textit{Mod~C} is fit to the data. Table \ref{tab:realC} and
Figure \ref{fig:realCfits} show the results of the fits. Figures
\ref{fig:realCgamma} and \ref{fig:realCNr} show that both $\gamma^i$
and $\Nr^i$ vary due to the loosening of hyper-prior constraints, but
$\Nr$ is again loosely constrained. We expect the estimates of
redshift to be unconstrained from simulations and find this in the
data as well (see Figure \ref{fig:realCz}).

Now, using the {\tt FITEXY} method, I find that the calibration
$N_{\rm cal}=51.17 \pm 0.02$ (compared to $\mean{\mu_{\Nr}} = 51.74$
with an HDI of $51.33 - 52.21$ for \textit{Mod~A}) obtained from
fitting the four GRBs with known redshift in the rest-frame with a
common $\gamma=1.72 \pm 0.02$ (compared to $\mean{\mu_\gamma }=1.49$
with an HDI of $1.35 - 1.67$ for \textit{Mod~A}). Table
\ref{tab:mleReal} reveals that while the method can estimate redshifts
precisely, known redshifts are reconstructed inaccurately. It was
determined that the errors of this method are likely underestimated
via simulations in Section \ref{sec:simA}.

Another option would be to use individual GRBs as calibration sources
and estimate known redshift. For each GRB with known redshift, I fit
its rest-frame GC to obtain $N_{\rm cal}$ and then proceed as before
using only one GRB as the calibration to estimate the known redshifts
of the other GRBs. Table \ref{tab:mleReal2} shows that the estimated
redshifts (columns) depend on which calibration GRB (rows) is
used. Regardless of which GRB is used as a calibration source, the
estimated redshifts are inaccurate. I conclude that the {\tt FITEXY}
method cannot be used to estimate redshifts.

\subsection{The Rest-Frame Golenetskii Relation}

I calculate the rest-frame Golenetskii relation for each model using
GRBs with known redshift following the procedure in Section
\ref{sec:simgol}. I exclude GRBs without known redshift due to the
inability to calculate rest-frame quantities when the redshift is
unconstrained as shown in Figure \ref{fig:golAll}. Figure
\ref{fig:realGol} shows the rest-frame correlation for all
models. There is little difference between the models'
predictions. Even with the hyper-parameter models, a difference in
the individual $\gamma^i$ and $\Nr^i$ is allowed. While $\gamma^i$ do
appear tightly distributed, the $\Nr^i$ can vary within an order of
magnitude.

\section{Discussion}

I have presented a hierarchical Bayesian model to test the ability of
the Golenetskii correlation (Equation \ref{eq:1}) to estimate the redshifts of
GRBs. The model incorporates all known variance in both the data and
assumptions and therefore provides a robust assessment of the claim
that accurate redshifts can be obtained. The model performs well with
simulated data. The data are generated under the assumptions of the
model, so it is expected, but fitting of the simulations allows for
determining the accuracy of the model. The model is able to predict
simulated redshifts very accurately as long as the rest-frame
normalization of the relations are all the same. If this assumption is
dropped, the model is unable to predict redshifts though it can still
recover rest-frame properties of GRBs with known redshift.

The method of using {\tt FITEXY} underestimates the errors on
predicted redshifts. More importantly, the method predicts inaccurate
redshifts if there is not a common $\Nr$ and $\gamma$ in the
rest-frame. Moreover, the inaccurate redshifts will have errors that
do not encompass the true redshift. Therefore, I find the method
unable to predict redshifts.

When I apply the three Bayesian models to the data, I find that the
redshift estimates are all unconstrained. Constrained estimates for
$\Nr^i$ for GRBs with known redshift and $\gamma^i$ for all GRBs are
recovered. Upon examining the inferred rest-frame Golenetskii
relations for GRBs with known redshift, I find that $\Nr^i$ are
tightly distributed but can vary within an order of magnitude. This is
most likely the reason that {\tt FITEXY} under-predicts known
redshift. To investigate this, I simulated 1000 GCs from
\textit{Mod~C} and fit them with {\tt FITEXY} as before. Then, the
distribution of the difference between estimated redshift and
simulated is examined as a function of the difference between $N_{\rm
  cal}$ and the simulated $\Nr$ (see Figure \ref{fig:ncal}). When
$N_{\rm cal}$ underestimates the true $\Nr$, then the redshift can be
under-predicted. As can be seen in Tables
\ref{tab:realA}-\ref{tab:realC}, $N_{\rm cal}$ always underestimates
$\Nr$ and therefore the {\tt FITEXY} method under predicts the known
redshifts in our sample.

This leads us to consider what happens when simulations from
\textit{Mod~C} (possibly resembling the real GRB sample) are fitted
but \textit{Mod~A}. We know from Section \ref{sec:simA} that
\textit{Mod~A} can perfectly estimate redshifts if it represents the
true generative model of the data. Therefore, I simulate data from
\textit{Mod~C} and fit it with \textit{Mod~A}. Figures
\ref{fig:simCANr} and \ref{fig:simCAz} show that, like the real data
in Section \ref{sec:real}, \textit{Mod~A} cannot constrain $\Nr$ and
subsequently redshift when there is not a common $\Nr$. Unfortunately,
\textit{Mod~C} which is applicable to the data, cannot constrain
redshifts. This strongly suggests that there is not a universal
Golenetskii correlation in the rest-frame.

The fact that all GRBs \textit{do not} share a common $\Nr$ is not
surprising. It is established that there are at least two observed
mechanisms occurring in GRBs: photospheric \citep{Ryde:2010} and
non-thermal emission \citep{Zhang:2015,Burgess:2015}. In both of these
cases, a variance in $\Nr$ is expected. I have shown that variance is
not negligible. The choice of spectral model also plays an crucial role in
determining the slope and normalization of GCs. The results of
\citet{Burgess:2014,Iyyani:2015tv} where a physical synchrotron model
was used to fit the spectra of the same GRBs used in this study
produced GCs with different slopes than what is found here with
empirical spectral models. This is due to the different curvature of
these models around the $\vFv$ peak which results in different $\Ep$'s
from the spectral fits.

Theoretical predictions for the rest-frame GC are in their
infancy. While weak predictions for $\gamma$ exist
\citep{Zhang:2002dr,Dermer:2004,Fan:2012,Bosnjak:2014}, exact
solutions have only begun to be formulated. This is due in part to the
stochastic nature of GRB lightcurves as well as the limited input
knowledge available for use in predicting the exact processes that
occur in GRB outflows. It is interesting to note that
\citet{Lopez-Camara:2014} predict that photospheric models can have
different normalizations in the rest-frame depending on viewing
angle. If physical spectra can be fit to the data corresponding to
such models, then the subsequently derived GCs could be used to predict
viewing angles of GRBs which presents an interesting new tool to study
GRB prompt emission mechanisms.

\section{Conclusions}

I conclude that the Golenetskii correlation does not possess a common
$\Nr$ which may have interesting implications for the outflow physics
of GRBs but precludes them from being used as redshift estimators
under the model posed herein. The Golenetskii correlation fails as a
standard candle without an additional predictor. This also precludes
the use of GCs to estimate cosmological parameters as the broad HDIs
found in this study would be folded into the errors on cosmological
parameters making their determination weak and unconstrained. GCs are
useful for discerning the physical emission mechanisms occurring in
GRBs and warrant dedicated study. A method similar to the one proposed
here can be used to study the observer- and/or rest-frame properties
of GCs provided the goal is to understand the properties intrinsic to
the GRB.

\section*{Acknowledgments}

I would like to thank Johannes Buchner, Michael Betancourt, and Bob
Carpenter for helpful discussion and direction on the probabilistic
model, {\tt Stan}, and font choice. Additionally, I would like to
thank Damien B\'{e}gu\'{e} and Felix Ryde for discussions on the
physics of GCs.

I additionally thank the Swedish National Infrastructure for Computing
(SNIC) and The PDC Center for High Performance Computing at the KTH
Royal Institute of Technology for computation time on the Tegn\'{e}r
cluster via grant PDC-2015-27.

This research made use of {\tt Astropy}, a community-developed core
Python package for Astronomy \citep{astropy} as well {\tt Matplotlib},
an open source Python graphics environment \citep{Hunter:2007} and
{\tt Seaborn} \citep{Waskom:2015} for plotting.
\bibliographystyle{mn2e}
\bibliography{bib}

\begin{thebibliography}{}

\bibitem[\protect\citeauthoryear{{Prior distributions for variance parameters
  in hierarchical models (comment on article by Browne and
  Draper)}}{sta}{2015}]{stan}
 2015, PyStan: the Python interface to Stan, Version 2.7.0

\bibitem[\protect\citeauthoryear{Adachi \& Kasai}{Adachi \&
  Kasai}{2012}]{Adachi:2012bn}
Adachi M.,  Kasai M.,  2012, Progress of Theoretical Physics, 127, 145

\bibitem[\protect\citeauthoryear{Amati et~al.,}{Amati
  et~al.}{2002}]{Amati:2002}
Amati L.,  et~al., 2002, 390, 81

\bibitem[\protect\citeauthoryear{Andreon}{Andreon}{2012}]{Andreon:2012dp}
Andreon S.,  2012, in , Astrostatistical Challenges for the New Astronomy.
Springer New York, New York, NY, pp 41--62

\bibitem[\protect\citeauthoryear{Andreon \& Hurn}{Andreon \&
  Hurn}{2013}]{Andreon:2013}
Andreon S.,  Hurn M.,  2013, Statistical Analysis and Data Mining: The {\ldots}

\bibitem[\protect\citeauthoryear{{Astropy Collaboration} et~al.,}{{Astropy
  Collaboration}  et~al.}{2013}]{astropy}
{Astropy Collaboration} et~al., 2013, A{\&}A, 558, A33

\bibitem[\protect\citeauthoryear{Band, Matteson \& Ford}{Band
  et~al.}{1993}]{Band:1993}
Band D.,  Matteson J.,    Ford L.,  1993, ApJ, 413, 281

\bibitem[\protect\citeauthoryear{B{\'e}gu{\'e} \& Pe'er}{B{\'e}gu{\'e} \&
  Pe'er}{2015}]{Begue:2015}
B{\'e}gu{\'e} D.,  Pe'er A.,  2015, ApJ, 802, 134

\bibitem[\protect\citeauthoryear{Beloborodov}{Beloborodov}{2010}]{Beloborodov:2010}
Beloborodov A.~M.,  2010, MNRAS, 407, 1033

\bibitem[\protect\citeauthoryear{Berger}{Berger}{2011}]{Berger:2011}
Berger E.,  2011, GCN Circular

\bibitem[\protect\citeauthoryear{Borgonovo \& Ryde}{Borgonovo \&
  Ryde}{2001}]{Borgonovo:2001}
Borgonovo L.,  Ryde F.,  2001, ApJ, 548, 770

\bibitem[\protect\citeauthoryear{Bo{\v s}njak \& Daigne}{Bo{\v s}njak \&
  Daigne}{2014}]{Bosnjak:2014}
Bo{\v s}njak {\v Z}.,  Daigne F.,  2014, A\&A, 568, A45

\bibitem[\protect\citeauthoryear{Burgess, B{\'e}gu{\'e}, Ryde, Omodei, Pe'er,
  Racusin \& Cucchiara}{Burgess et~al.}{2015}]{Burgess:2015}
Burgess J.~M.,  B{\'e}gu{\'e} D.,  Ryde F.,  Omodei N.,  Pe'er A.,  Racusin
  J.~L.,    Cucchiara A.,  2015, arXiv.org

\bibitem[\protect\citeauthoryear{Burgess et~al.,}{Burgess
  et~al.}{2014}]{Burgess:2014}
Burgess J.~M.,  et~al., 2014, ApJ, 784, 17

\bibitem[\protect\citeauthoryear{de Souza, Hilbe, Buelens, Riggs, Cameron,
  Ishida, Chies-Santos \& Killedar}{de~Souza et~al.}{2015}]{deSouza:2015gk}
de Souza R.~S.,  Hilbe J.~M.,  Buelens B.,  Riggs J.~D.,  Cameron E.,  Ishida
  E. E.~O.,  Chies-Santos A.~L.,    Killedar M.,  2015, Monthly Notices of the
  Royal Astronomical Society, 453, 1928

\bibitem[\protect\citeauthoryear{Dellaportas \& Stephens}{Dellaportas \&
  Stephens}{1995}]{Dellaportas:1995}
Dellaportas P.,  Stephens D.~A.,  1995, Biometrics, 51, 1085

\bibitem[\protect\citeauthoryear{Dermer}{Dermer}{2004}]{Dermer:2004}
Dermer C.~D.,  2004, ApJ, 614, 284

\bibitem[\protect\citeauthoryear{Fan, Wei, Zhang \& Zhang}{Fan
  et~al.}{2012}]{Fan:2012}
Fan Y.-Z.,  Wei D.-M.,  Zhang F.-W.,    Zhang B.-B.,  2012, ApJL, 755, L6

\bibitem[\protect\citeauthoryear{Gelman}{Gelman}{2006}]{Gelman:2006di}
Gelman A.,  2006, Bayesian analysis, 1, 515

\bibitem[\protect\citeauthoryear{Ghirlanda, Nava \& Ghisellini}{Ghirlanda
  et~al.}{2010}]{Ghirlanda:2010}
Ghirlanda G.,  Nava L.,    Ghisellini G.,  2010, A\&A, 511, 43

\bibitem[\protect\citeauthoryear{Giannios}{Giannios}{2006}]{Giannios:2006}
Giannios D.,  2006, A{\&}A, 457, 763

\bibitem[\protect\citeauthoryear{Golenetskii, Mazets, Aptekar \&
  Ilinskii}{Golenetskii et~al.}{1983}]{Golenetskii:1983}
Golenetskii S.~V.,  Mazets E.~P.,  Aptekar R.~L.,    Ilinskii V.~N.,  1983,
  Nature, 306, 451

\bibitem[\protect\citeauthoryear{Greiner}{Greiner}{2011}]{Griener:2011}
Greiner J.,  2011, GCN Circular

\bibitem[\protect\citeauthoryear{Guiriec et~al.,}{Guiriec
  et~al.}{2013}]{Guiriec:2013hl}
Guiriec S.,  et~al., 2013, ApJ, 770, 32

\bibitem[\protect\citeauthoryear{Guiriec et~al.,}{Guiriec
  et~al.}{2015}]{Guiriec:2015}
Guiriec S.,  et~al., 2015, ApJ

\bibitem[\protect\citeauthoryear{Hunter}{Hunter}{2007}]{Hunter:2007}
Hunter J.~D.,  2007, Computing In Science \& Engineering, 9, 90

\bibitem[\protect\citeauthoryear{Iyyani, Ryde, Burgess, Pe'er \&
  egu{\'e}}{Iyyani et~al.}{2015}]{Iyyani:2015tv}
Iyyani S.,  Ryde F.,  Burgess J.~M.,  Pe'er A.,    egu{\'e} D.~B.,  2015,
  arXiv.org

\bibitem[\protect\citeauthoryear{Kargatis}{Kargatis}{1994}]{Kargatis:1994}
Kargatis V.,  1994, AAS, 185, 1508

\bibitem[\protect\citeauthoryear{Kelly}{Kelly}{2007}]{Kelly:2007}
Kelly B.~C.,  2007, ApJ, 665, 1489

\bibitem[\protect\citeauthoryear{Kocevski}{Kocevski}{2012}]{Kocevski:2012}
Kocevski D.,  2012, ApJ, 747, 146

\bibitem[\protect\citeauthoryear{L{\'o}pez-C{\'a}mara, Morsony \&
  Lazzati}{L{\'o}pez-C{\'a}mara et~al.}{2014}]{Lopez-Camara:2014}
L{\'o}pez-C{\'a}mara D.,  Morsony B.~J.,    Lazzati D.,  2014, MNRAS, 442, 2202

\bibitem[\protect\citeauthoryear{Lu, Hou \& Liang}{Lu et~al.}{2010}]{Lu:2010bj}
Lu R.~J.,  Hou S.~J.,    Liang E.-W.,  2010, ApJ, 720, 1146

\bibitem[\protect\citeauthoryear{Lundman, Pe'er \& Ryde}{Lundman
  et~al.}{2014}]{Lundman:2014}
Lundman C.,  Pe'er A.,    Ryde F.,  2014, MNRAS, 440, 3292

\bibitem[\protect\citeauthoryear{Mandel, Narayan \& Kirshner}{Mandel
  et~al.}{2011}]{Mandel:2011}
Mandel K.~S.,  Narayan G.,    Kirshner R.~P.,  2011, ApJ, 731, 120

\bibitem[\protect\citeauthoryear{March, Trotta, Berkes, Starkman \&
  Vaudrevange}{March et~al.}{2011}]{March:2011jx}
March M.~C.,  Trotta R.,  Berkes P.,  Starkman G.~D.,    Vaudrevange P.~M.,
  2011, MNRAS, 418, 2308

\bibitem[\protect\citeauthoryear{Massaro, Cutini, Conciatore \&
  Tramacere}{Massaro et~al.}{2007}]{Massaro:2007}
Massaro F.,  Cutini S.,  Conciatore M.~L.,    Tramacere A.,  2007, arXiv.org,
  pp 84--87

\bibitem[\protect\citeauthoryear{Pe'er, Meszaros \& Rees}{Pe'er
  et~al.}{2005}]{Peer:2005}
Pe'er A.,  Meszaros P.,    Rees M.~J.,  2005, ApJ, 635, 476

\bibitem[\protect\citeauthoryear{Preece, Burgess et~al.,}{Preece
  et~al.}{2014}]{Preece:2014}
Preece R.,  Burgess J.~M.,    et~al., 2014, Science, 343, 51

\bibitem[\protect\citeauthoryear{Press}{Press}{2007}]{Press:2007}
Press W.~H.,  2007, {Numerical Recipes 3rd Edition}.
The Art of Scientific Computing, Cambridge University Press

\bibitem[\protect\citeauthoryear{Ryde, Bjornsson, Kaneko, Meszaros, Preece \&
  Battelino}{Ryde et~al.}{2006}]{Ryde:2006}
Ryde F.,  Bjornsson C.-I.,  Kaneko Y.,  Meszaros P.,  Preece R.,    Battelino
  M.,  2006, ApJ, 652, 1400

\bibitem[\protect\citeauthoryear{Ryde et~al.,}{Ryde  et~al.}{2010}]{Ryde:2010}
Ryde F.,  et~al., 2010, ApJL, 709, L172

\bibitem[\protect\citeauthoryear{Sanders et~al.,}{Sanders
  et~al.}{2015}]{Sanders:2015}
Sanders N.~E.,  et~al., 2015, ApJ, 799, 208

\bibitem[\protect\citeauthoryear{Waskom et~al.,}{Waskom
  et~al.}{2015}]{Waskom:2015}
Waskom M.,  et~al.,, 2015, seaborn: v0.6.0 (June 2015)

\bibitem[\protect\citeauthoryear{Yonetoku, Murakami \& Nakamura}{Yonetoku
  et~al.}{2004}]{Yonetoku:2004}
Yonetoku D.,  Murakami T.,    Nakamura T.,  2004, ApJ

\bibitem[\protect\citeauthoryear{Zhang \& Meszaros}{Zhang \&
  Meszaros}{2002}]{Zhang:2002dr}
Zhang B.,  Meszaros P.,  2002, ApJ

\bibitem[\protect\citeauthoryear{Zhang, Uhm, Connaughton, Briggs \&
  Zhang}{Zhang et~al.}{2015}]{Zhang:2015}
Zhang B.-B.,  Uhm Z.~L.,  Connaughton V.,  Briggs M.~S.,    Zhang B.,  2015,
  arXiv.org, p.~5858

\bibitem[\protect\citeauthoryear{Zhang \& Chen}{Zhang \&
  Chen}{2012}]{Zhang:2012fq}
Zhang Z.~B.,  Chen D.~Y.,  2012, ApJ, 755, 55

\end{thebibliography}

\clearpage



\begin{figure}
  \centering
  \includegraphics{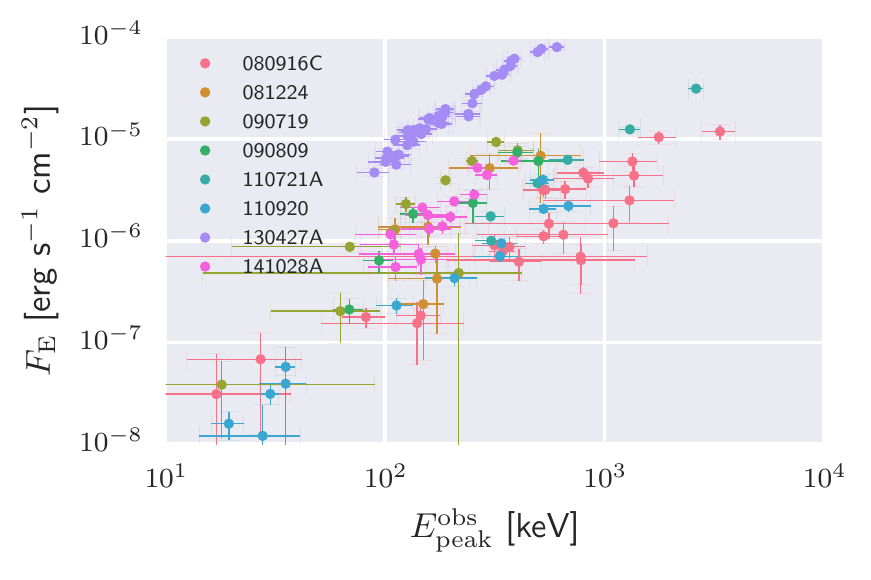}
  \caption{The observer-frame GCs for real GRBs in our sample.}
  \label{fig:realdata}
\end{figure}


\begin{figure}
  \centering
  \includegraphics{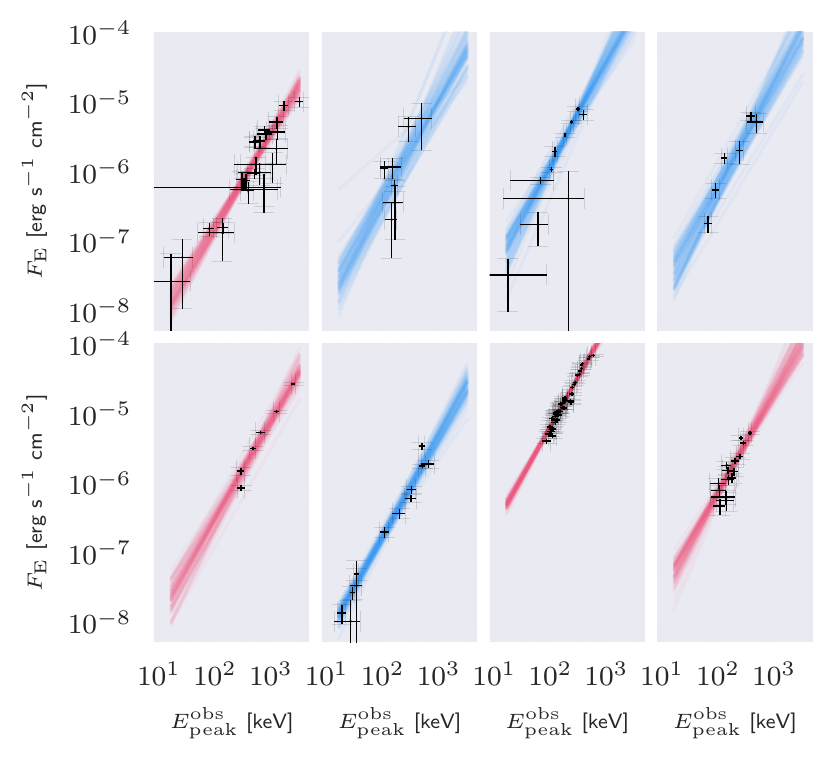}
  \caption{The reconstructed observer-frame GCs for real data fit with
    \textit{Mod~A}. The \textit{red} lines map out the posterior of
    known redshift GRBs while \textit{blue} maps out the posterior of
    unknown redshift GRBs. The plots from top-left to bottom right
    display GRBs 080916C, 081224A, 090719A, 090809A, 110721A, 110920A,
    130427A, and 141028A respectively.}
  \label{fig:realAfits}
\end{figure}

\begin{figure}
  \centering
  \includegraphics{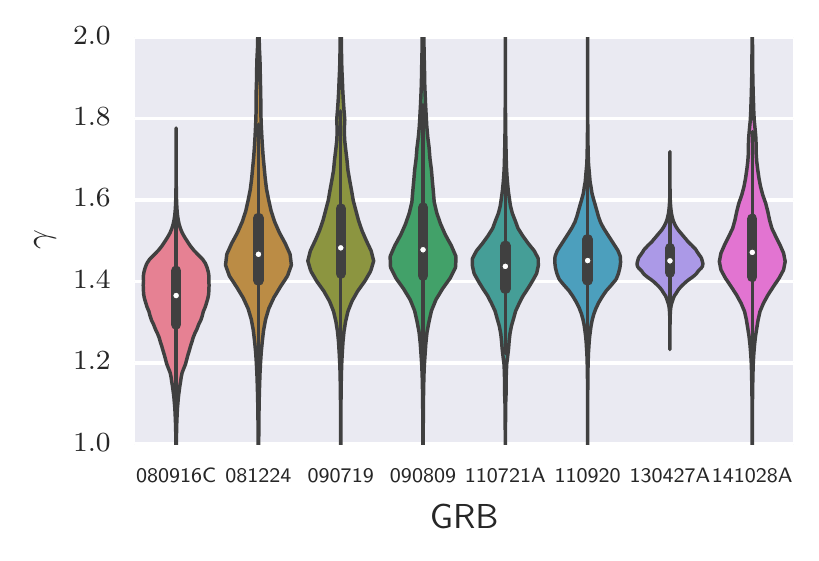}
  \caption{The estimated distributions of $\gamma^i$ for real data fit
    with \textit{Mod~A}.}
  \label{fig:realAgamma}
\end{figure}

\begin{figure}
  \centering
  \includegraphics{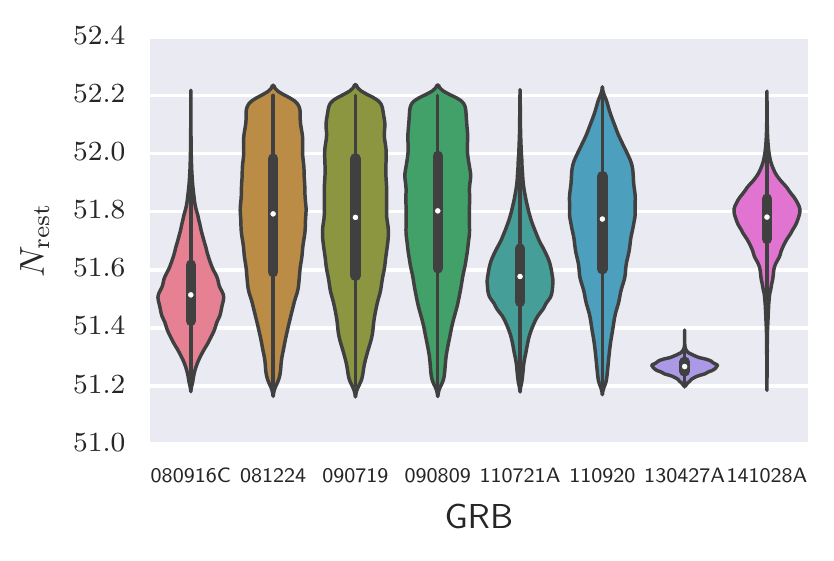}
  \caption{The estimated distributions of $\Nr^i$ for real data fit
    with \textit{Mod~A}.}
  \label{fig:realANr}
\end{figure}

\begin{figure}
  \centering
  \includegraphics{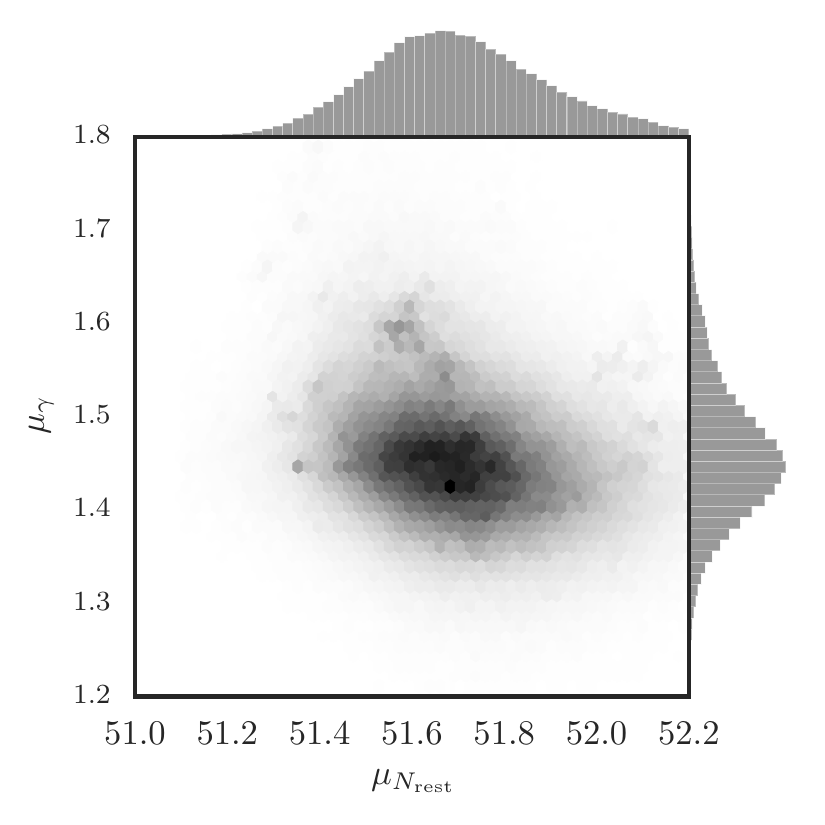}
  \caption{The distributions of hyper-parameters for $\Nr^i$ and
    $\gamma^i$ for real data fit with \textit{Mod~A}.}
  \label{fig:realAmu}
\end{figure}

\begin{figure}
  \centering
  \includegraphics{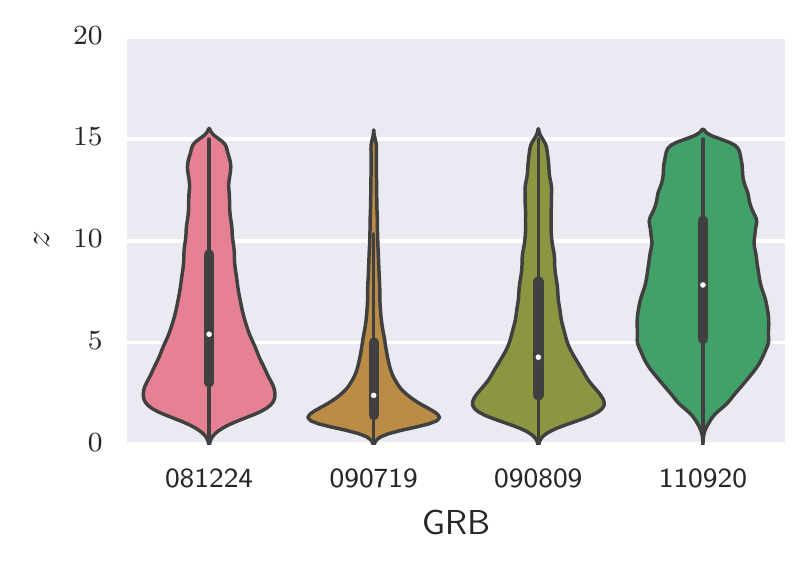}
  \caption{The estimated distributions of unknown redshifts for real
    data fit with \textit{Mod~A}}
  \label{fig:realAz}
\end{figure}


\begin{figure}
  \centering
  \includegraphics{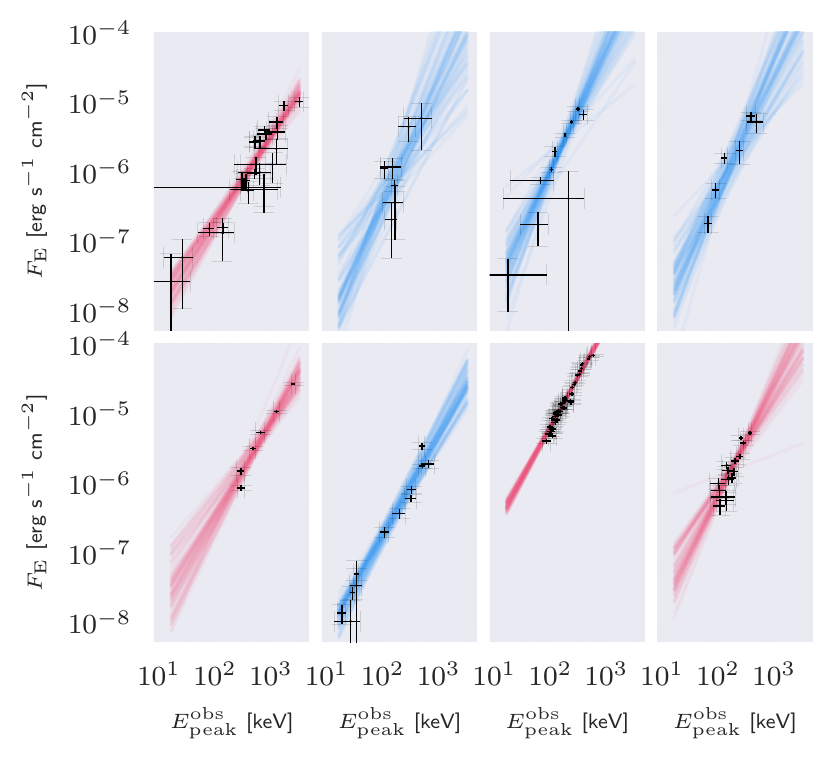}
  \caption{Same as Figure \ref{fig:realAfits} but for \textit{Mod~B}.}
  \label{fig:realBfits}
\end{figure}

\begin{figure}
  \centering
  \includegraphics{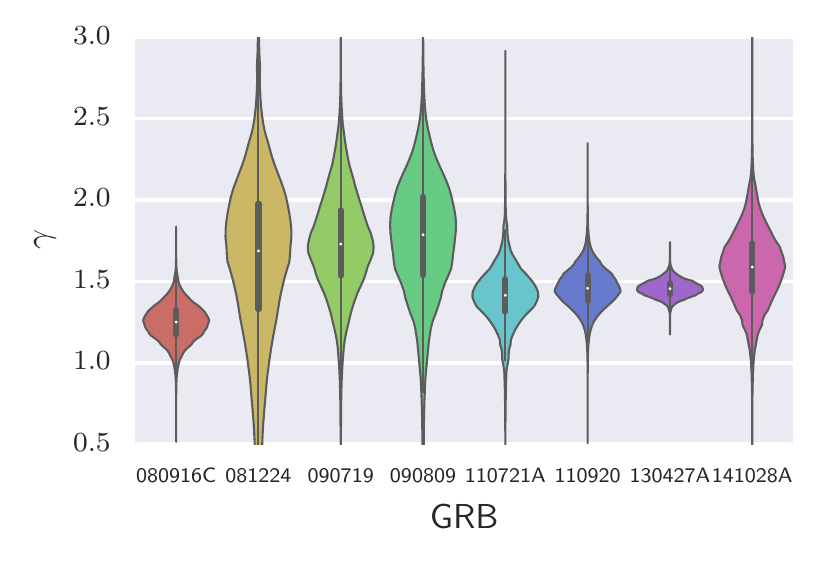}
  \caption{Same as Figure \ref{fig:realAgamma} but for \textit{Mod~B}.}
  \label{fig:realBgamma}
\end{figure}

\begin{figure}
  \centering
  \includegraphics{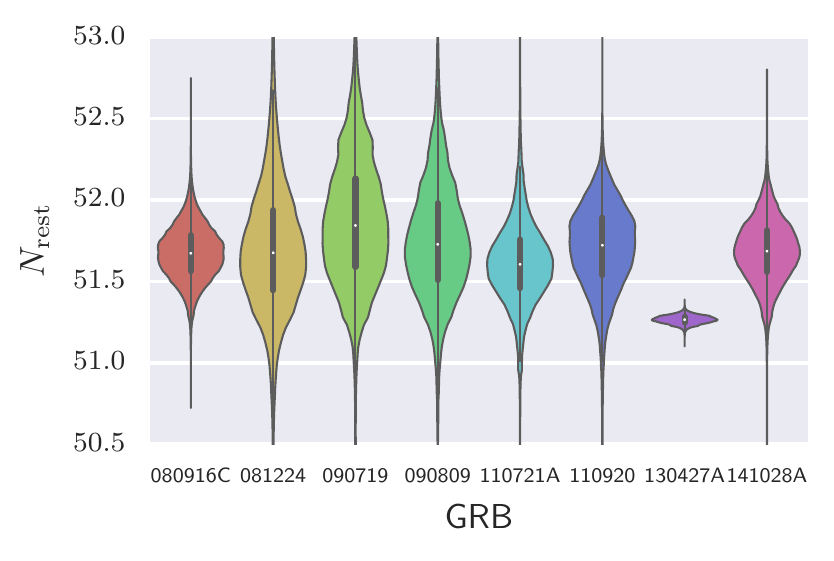}
  \caption{Same as Figure \ref{fig:realANr} but for \textit{Mod~B}.}
  \label{fig:realBNr}
\end{figure}

\begin{figure}
  \centering
  \includegraphics{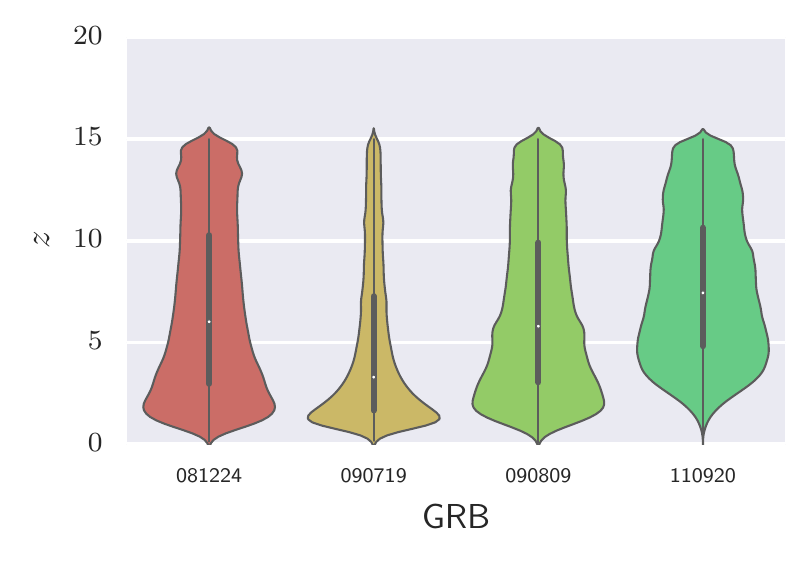}
  \caption{Same as Figure \ref{fig:realAz} but for \textit{Mod~B}.}
  \label{fig:realBz}
\end{figure}


\begin{figure}
  \centering
  \includegraphics{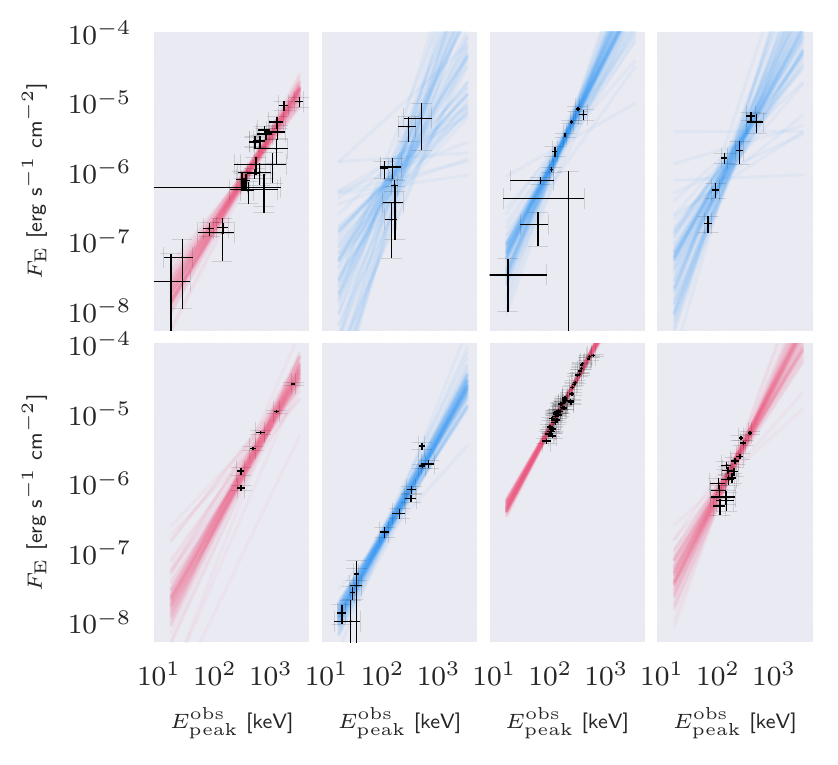}
 \caption{Same as Figure \ref{fig:realAfits} but for \textit{Mod~C}.}
  \label{fig:realCfits}
\end{figure}

\begin{figure}
  \centering
  \includegraphics{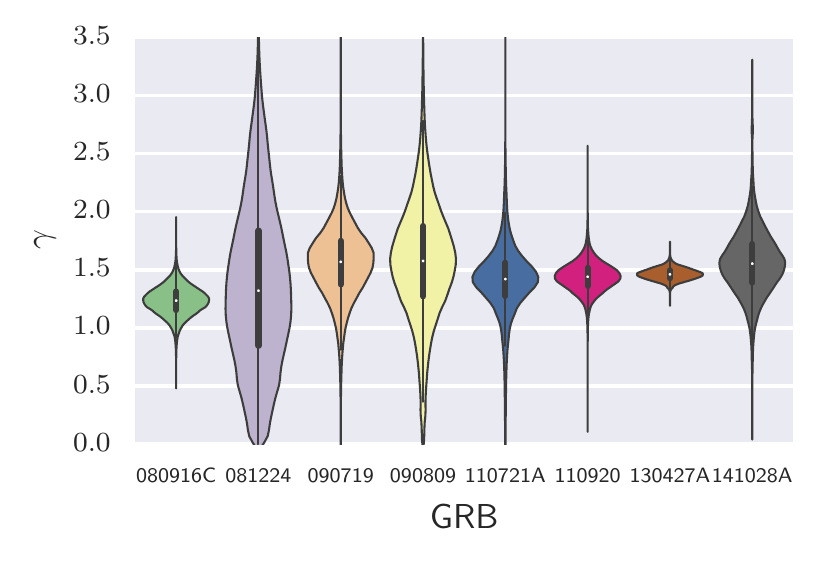}
 \caption{Same as Figure \ref{fig:realAgamma} but for \textit{Mod~C}.}
  \label{fig:realCgamma}
\end{figure}

\begin{figure}
  \centering
  \includegraphics{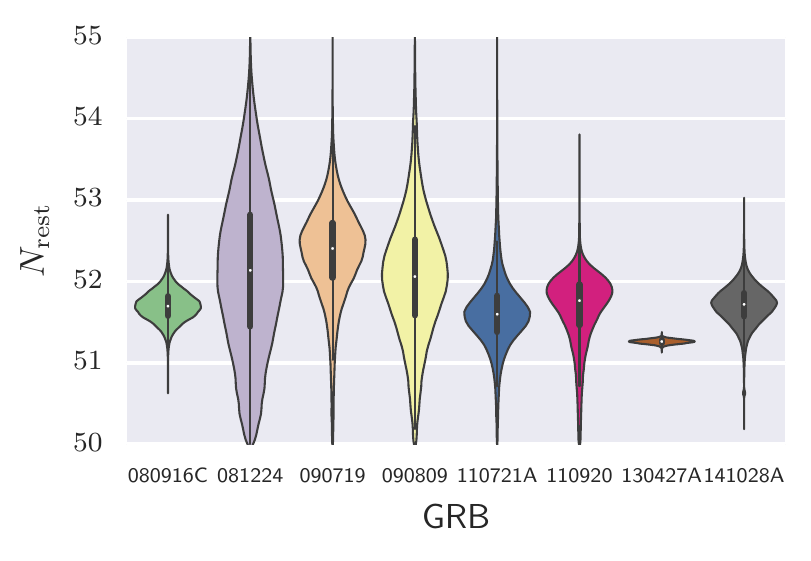}
  \caption{Same as Figure \ref{fig:realANr} but for \textit{Mod~C}.}
  \label{fig:realCNr}
\end{figure}

\begin{figure}
  \centering
  \includegraphics{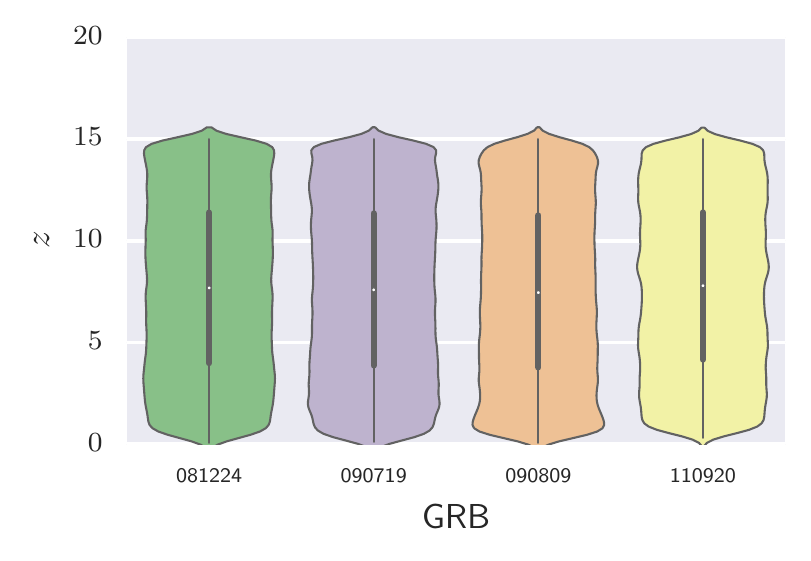}
 \caption{Same as Figure \ref{fig:realAz} but for \textit{Mod~C}.}
  \label{fig:realCz}
\end{figure}

\begin{figure}
  \centering
  \includegraphics{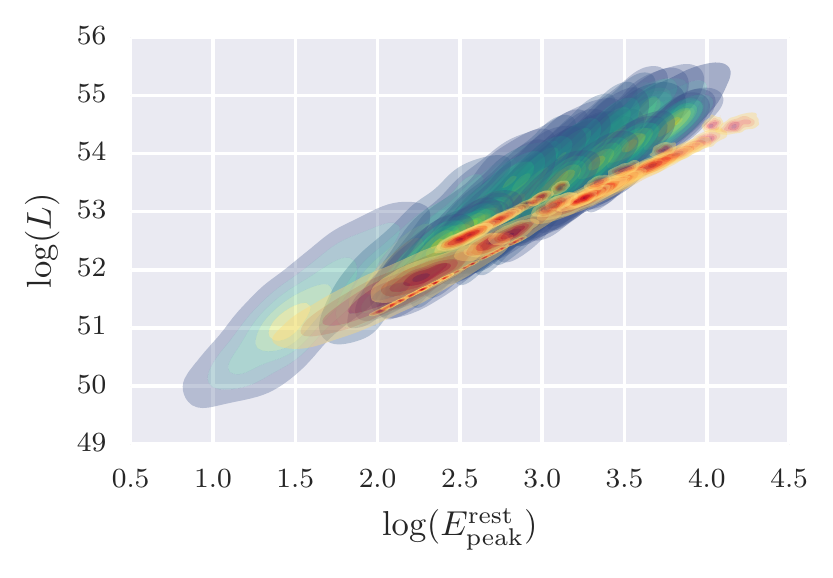}
  \caption{The rest-frame GC for real GRBs fit with \textit{Mod A} demonstrating the difference in constraints for GRBs with known redshift (yellow-red) and without known redshift (blue-green).}
  \label{fig:golAll}
\end{figure}
\clearpage

\begin{figure}
  \centering

\subfigure[]{\includegraphics{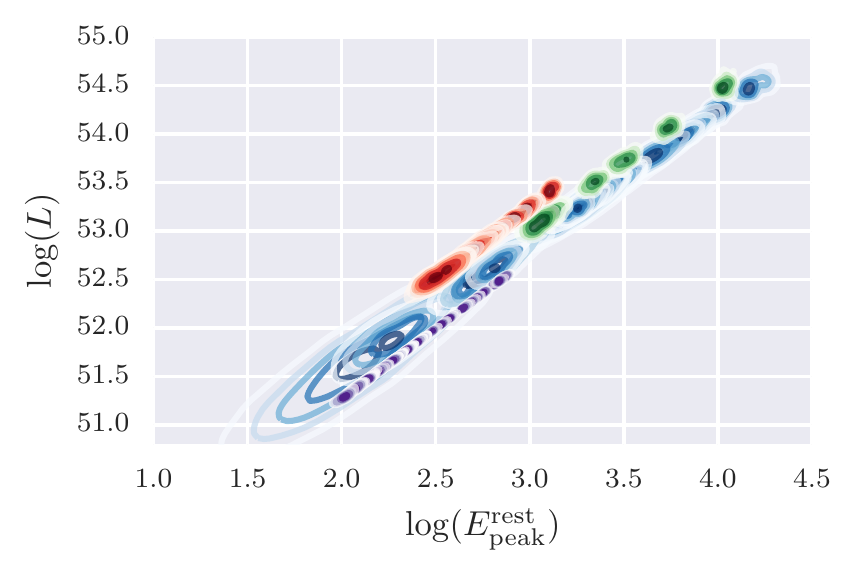}}
\subfigure[]{\includegraphics{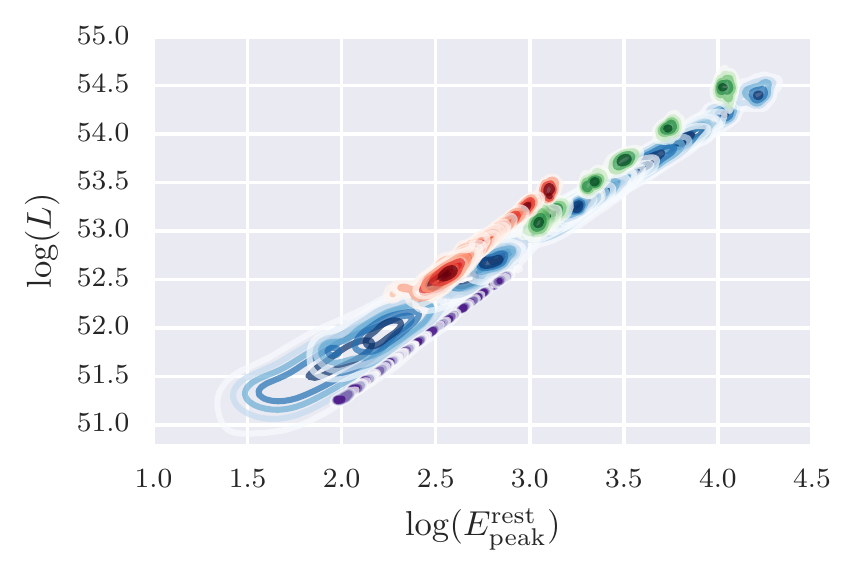}}
\subfigure[]{\includegraphics{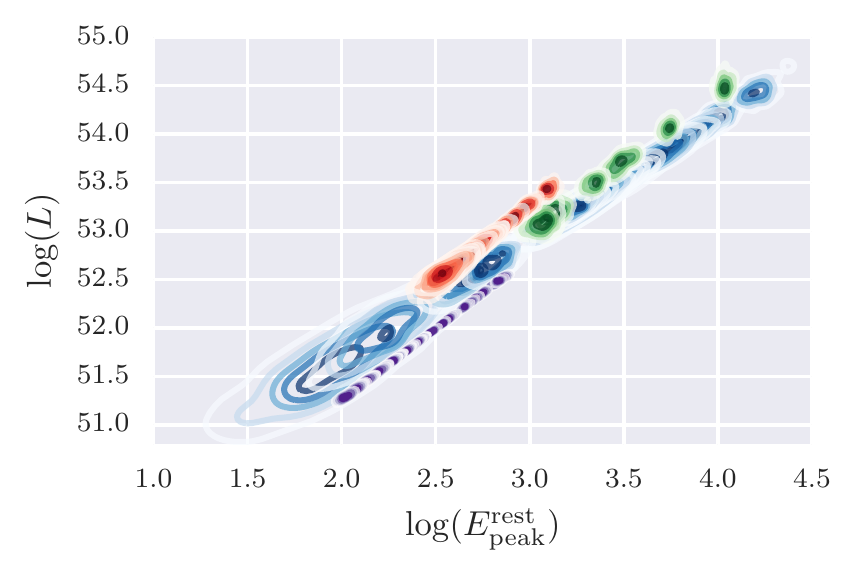}}
 
\caption{From top to bottom, the predicted rest-frame Golenetskii
    correlations for GRBs with known redshift from \textit{Mod
      A}-\textit{Mod~C} respectively. GRBs 080916C (blue), 110721A
    (red), 130427A (purple), and 141028A (purple) are displayed. Take
    note of the differences in amplitude of the GCs even though all
    are fit with different models.}
  \label{fig:realGol}
\end{figure}

\begin{figure}
  \centering
\includegraphics{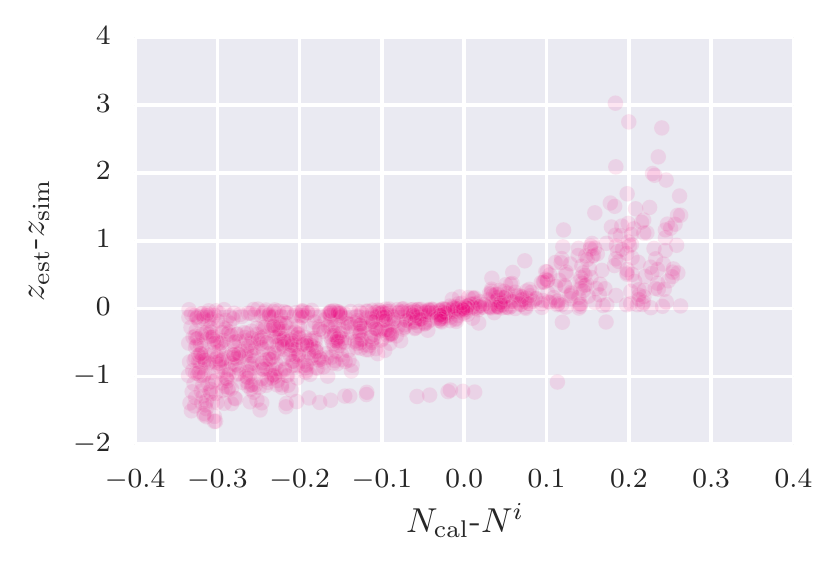}
  \caption{The distribution of miscalculated redshifts as a function
    of the distance of the simulated $\Nr$ from $N_{\rm cal}$ for fits
    performed with {\tt FITEXY} showing a clear relation of
    underestimated redshift when the true $\Nr$ is greater than
    $N_{\rm cal}$ as is observed in the real data.}
\label{fig:ncal}
\end{figure}

\begin{figure}
  \centering
  \includegraphics{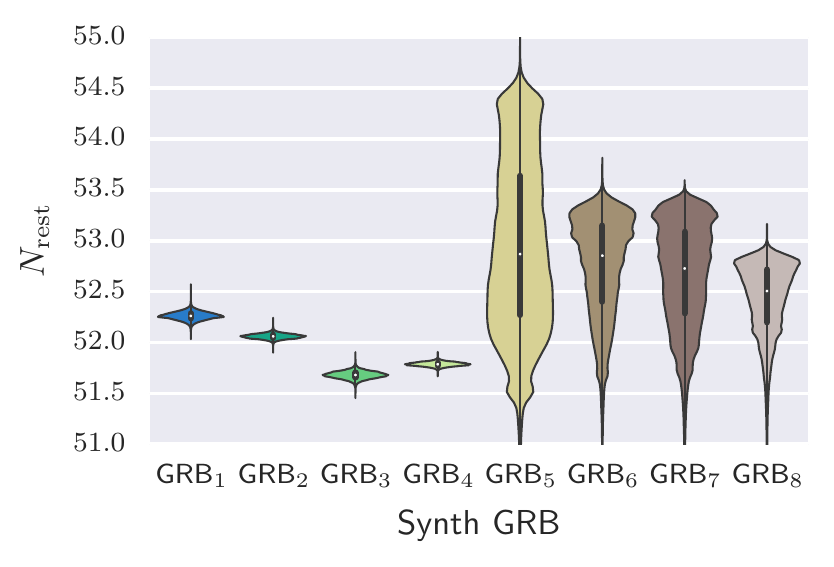}
  \caption{The $\Nr^i$ for GCs simualted from \textit{Mod~C} but fit
    with \textit{Mod~A}. GRBs with unknown redshift have unconstrained
    $\Nr$ similar to what is observered in real data.}
  \label{fig:simCANr}
\end{figure}

\begin{figure}
  \centering
  \includegraphics{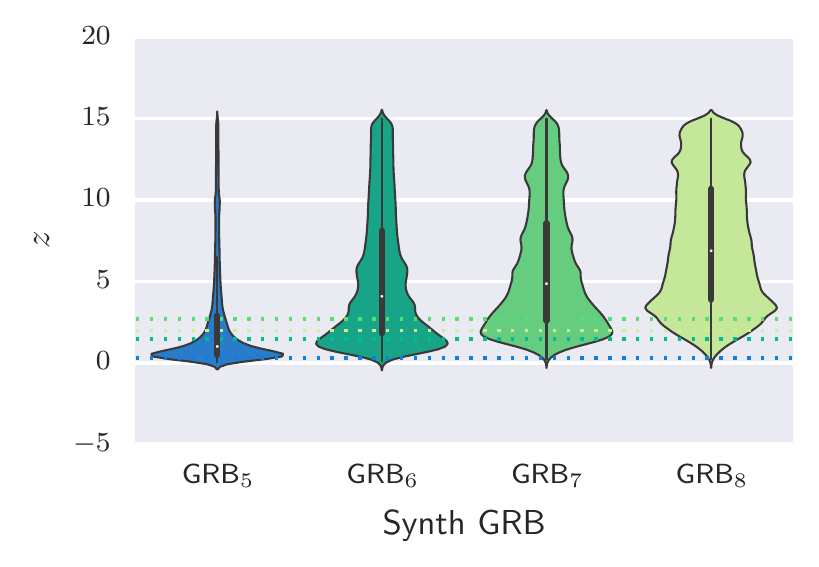}
  \caption{The predicted $z^i$ for GCs simualted from \textit{Mod~C}
    but fit with \textit{Mod~A}. GRBs with unknown redshift have
    unconstrained $z$ similar to what is observered in real data
    (compare to Figure \ref{fig:realAz}).}
  \label{fig:simCAz}
\end{figure}

\clearpage

\begin{deluxetable}{cccccccccc}

\tablecolumns{10}
\tablewidth{0pt}
\tablecaption{Fit Results of \textit{Mod~A} on Simulations}
\tablehead{\colhead{Synth. GRB} & \colhead{$z$} & \colhead{$\mean{z_{\rm est}}$} & \colhead{$z{\rm est}$ 95\% HDI} & \colhead{$N_{\rm rest}$} & \colhead{$\mean{N_{\rm rest, est}}$} & \colhead{$N_{\rm rest, est}$ 95\% HDI} & \colhead{$\gamma$} & \colhead{$\mean{\gamma_{\rm est}}$} & \colhead{$\gamma_{\rm est}$ 95\% HDI}}
\startdata
GRB$_{0}$ & 0.60 &\nodata& \nodata & 52.00 & 52.02 & 51.99 - 52.05 & 1.50 & 1.48 & 1.44 - 1.52 \\
GRB$_{1}$ & 2.48 &\nodata& \nodata & 52.00 & 52.02 & 51.99 - 52.04 & 1.50 & 1.48 & 1.45 - 1.52 \\
GRB$_{2}$ & 2.06 &\nodata& \nodata & 52.00 & 52.02 & 51.98 - 52.05 & 1.50 & 1.48 & 1.44 - 1.52 \\
GRB$_{3}$ & 1.85 &\nodata& \nodata & 52.00 & 52.01 & 51.99 - 52.04 & 1.50 & 1.48 & 1.44 - 1.51 \\
GRB$_{4}$ & 0.29 & 0.32 & 0.26 - 0.34 & 52.00 & 52.04 & 51.91 - 52.12 & 1.50 & 1.48 & 1.44 - 1.52 \\
GRB$_{5}$ & 1.47 & 1.57 & 1.27 - 1.75 & 52.00 & 52.03 & 51.92 - 52.14 & 1.50 & 1.47 & 1.43 - 1.51 \\
GRB$_{6}$ & 2.70 & 2.86 & 2.27 - 3.46 & 52.00 & 52.03 & 51.94 - 52.16 & 1.50 & 1.48 & 1.44 - 1.52 \\
GRB$_{7}$ & 1.99 & 2.13 & 1.74 - 2.43 & 52.00 & 52.03 & 51.91 - 52.12 & 1.50 & 1.49 & 1.46 - 1.52 \\

\enddata
\label{tab:simA}
\end{deluxetable}

\tablecaption{Fit Results from {\tt FITEXY} on Simulated Data from \textit{Mod~A}}
\begin{deluxetable}{ccccc}
\tablecolumns{5}
\tablewidth{0pt}
\tablehead{\colhead{Synth. GRB} & \colhead{$z$} & \colhead{$z_{\rm est}$} & \colhead{$\gamma$} & \colhead{$\gamma_{\rm est}$}}
\startdata
GRB$_{0}$ & 0.60 & $0.57 \pm 0.04$ & $1.46 \pm 0.05$ & 1.50 \\
GRB$_{1}$ & 2.48 & $2.45 \pm 0.14$ & $1.49 \pm 0.03$ & 1.50 \\
GRB$_{2}$ & 2.06 & $2.11 \pm 0.23$ & $1.51 \pm 0.06$ & 1.50 \\
GRB$_{3}$ & 1.85 & $1.85 \pm 0.06$ & $1.48 \pm 0.02$ & 1.50 \\
GRB$_{4}$ & 0.29 & $0.29 \pm 0.01$ & $1.47 \pm 0.03$ & 1.50 \\
GRB$_{5}$ & 1.47 & $1.45 \pm 0.05$ & $1.48 \pm 0.02$ & 1.50 \\
GRB$_{6}$ & 2.70 & $2.70 \pm 0.15$ & $1.50 \pm 0.03$ & 1.50 \\
GRB$_{7}$ & 1.99 & $2.02 \pm 0.06$ & $1.51 \pm 0.02$ & 1.50 \\
\enddata

\label{tab:mleA}
\end{deluxetable}

\begin{deluxetable}{cccccccccc}
\tablecaption{Fit Results of \textit{Mod~B} on Simulations}
\tablecolumns{10}
\tablewidth{0pt}
\tablehead{\colhead{Synth. GRB} & \colhead{$z$} & \colhead{$\mean{z_{\rm est}}$} & \colhead{$z{\rm est}$ 95\% HDI} & \colhead{$N_{\rm rest}$} & \colhead{$\mean{N_{\rm rest, est}}$} & \colhead{$N_{\rm rest, est}$ 95\% HDI} & \colhead{$\gamma$} & \colhead{$\mean{\gamma_{\rm est}}$} & \colhead{$\gamma_{\rm est}$ 95\% HDI}}
\startdata
GRB$_{0}$ & 0.60 &\nodata& \nodata & 52.00 & 52.02 & 51.95 - 52.11 & 1.87 & 1.83 & 1.71 - 1.93 \\
GRB$_{1}$ & 2.48 &\nodata& \nodata & 52.00 & 52.01 & 51.98 - 52.05 & 1.43 & 1.40 & 1.34 - 1.46 \\
GRB$_{2}$ & 2.06 &\nodata& \nodata & 52.00 & 52.01 & 51.97 - 52.06 & 1.79 & 1.75 & 1.68 - 1.83 \\
GRB$_{3}$ & 1.85 &\nodata& \nodata & 52.00 & 52.02 & 51.98 - 52.05 & 1.36 & 1.34 & 1.27 - 1.41 \\
GRB$_{4}$ & 0.29 & 0.35 & 0.19 - 0.55 & 52.00 & 52.08 & 51.76 - 52.65 & 1.20 & 1.20 & 1.16 - 1.25 \\
GRB$_{5}$ & 1.47 & 1.69 & 0.96 - 2.35 & 52.00 & 52.05 & 51.71 - 52.34 & 1.40 & 1.37 & 1.33 - 1.42 \\
GRB$_{6}$ & 2.70 & 3.05 & 2.22 - 5.99 & 52.00 & 52.06 & 51.92 - 52.55 & 1.26 & 1.20 & 1.13 - 1.26 \\
GRB$_{7}$ & 1.99 & 2.32 & 1.56 - 4.64 & 52.00 & 52.06 & 51.89 - 52.50 & 1.54 & 1.52 & 1.39 - 1.64 \\
\enddata
\label{tab:simB}
\end{deluxetable}

\begin{deluxetable}{cccccc}
\tablecolumns{6}
\tablewidth{0pt}
\tablecaption{Fit Results from {\tt FITEXY} on Simulated Data from \textit{Mod~B}}
\tablehead{\colhead{Synth. GRB} & \colhead{$z$} & \colhead{$z_{\rm est}$} & \colhead{$N$} & \colhead{$\gamma$} & \colhead{$\gamma_{\rm est}$}}
\startdata
GRB$_{0}$ & 0.60 & $0.54 \pm 0.05$ & 52.00 & 1.87 & $1.87 \pm 0.06$ \\
GRB$_{1}$ & 2.48 & $2.16 \pm 0.11$ & 52.00 & 1.43 & $1.41 \pm 0.03$ \\
GRB$_{2}$ & 2.06 & $1.81 \pm 0.06$ & 52.00 & 1.79 & $1.77 \pm 0.02$ \\
GRB$_{3}$ & 1.85 & $1.62 \pm 0.07$ & 52.00 & 1.36 & $1.34 \pm 0.03$ \\
GRB$_{4}$ & 0.29 & $0.26 \pm 0.01$ & 52.00 & 1.20 & $1.21 \pm 0.02$ \\
GRB$_{5}$ & 1.47 & $1.29 \pm 0.03$ & 52.00 & 1.40 & $1.38 \pm 0.01$ \\
GRB$_{6}$ & 2.70 & $2.37 \pm 0.06$ & 52.00 & 1.26 & $1.24 \pm 0.02$ \\
GRB$_{7}$ & 1.99 & $1.73 \pm 0.11$ & 52.00 & 1.54 & $1.52 \pm 0.04$ \\
\enddata
\label{tab:mleB}
\end{deluxetable}

\clearpage

\begin{deluxetable}{cccccccccc}
\tablecaption{Fit Results of \textit{Mod~C} on Simulations}
\tablecolumns{10}
\tablewidth{0pt}
\tablehead{\colhead{Synth. GRB} & \colhead{$z$} & \colhead{$\mean{z_{\rm est}}$} & \colhead{$z{\rm est}$ 95\% HDI} & \colhead{$N_{\rm rest}$} & \colhead{$\mean{N_{\rm rest, est}}$} & \colhead{$N_{\rm rest, est}$ 95\% HDI} & \colhead{$\gamma$} & \colhead{$\mean{\gamma_{\rm est}}$} & \colhead{$\gamma_{\rm est}$ 95\% HDI}}
\startdata
GRB$_{0}$ & 0.60 &\nodata& \nodata & 52.26 & 52.26 & 52.18 - 52.34 & 1.32 & 1.32 & 1.22 - 1.42 \\
GRB$_{1}$ & 2.48 &\nodata& \nodata & 52.06 & 52.06 & 52.00 - 52.10 & 1.40 & 1.40 & 1.33 - 1.49 \\
GRB$_{2}$ & 2.06 &\nodata& \nodata & 51.72 & 51.69 & 51.61 - 51.76 & 1.06 & 1.14 & 1.03 - 1.26 \\
GRB$_{3}$ & 1.85 &\nodata& \nodata & 51.77 & 51.79 & 51.74 - 51.83 & 1.31 & 1.27 & 1.21 - 1.33 \\
GRB$_{4}$ & 0.29 & 7.43 & 0.04 - 14.18 & 51.75 & 53.96 & 52.56 - 54.71 & 1.35 & 1.27 & 1.16 - 1.38 \\
GRB$_{5}$ & 1.47 & 7.66 & 0.52 - 14.61 & 52.24 & 52.95 & 51.87 - 53.48 & 1.63 & 1.57 & 1.48 - 1.66 \\
GRB$_{6}$ & 2.70 & 7.79 & 1.04 - 15.00 & 52.30 & 52.84 & 51.54 - 53.47 & 1.02 & 0.99 & 0.94 - 1.04 \\
GRB$_{7}$ & 1.99 & 7.48 & 0.65 - 14.62 & 51.72 & 52.37 & 51.08 - 52.92 & 1.44 & 1.32 & 1.24 - 1.40 \\
\enddata
\label{tab:simC}
\end{deluxetable}

\begin{deluxetable}{cccccc}
\tablecaption{Fit Results from {\tt FITEXY} on Simulated Data from \textit{Mod~C}}
\tablecolumns{6}
\tablewidth{0pt}
\tablehead{\colhead{Synth. GRB} & \colhead{$z$} & \colhead{$z_{\rm est}$} & \colhead{$N$} & \colhead{$\gamma$} & \colhead{$\gamma_{\rm est}$}}
\startdata
GRB$_{0}$ & 0.60 & $0.26 \pm 0.02$ & 52.26 & 1.32 & $1.32 \pm 0.04$ \\
GRB$_{1}$ & 2.48 & $1.14 \pm 0.04$ & 52.06 & 1.40 & $1.40 \pm 0.02$ \\
GRB$_{2}$ & 2.06 & $1.65 \pm 0.11$ & 51.72 & 1.06 & $1.15 \pm 0.05$ \\
GRB$_{3}$ & 1.85 & $1.29 \pm 0.05$ & 51.77 & 1.31 & $1.28 \pm 0.03$ \\
GRB$_{4}$ & 0.29 & $0.21 \pm 0.01$ & 51.75 & 1.35 & $1.29 \pm 0.03$ \\
GRB$_{5}$ & 1.47 & $0.55 \pm 0.03$ & 52.24 & 1.63 & $1.59 \pm 0.04$ \\
GRB$_{6}$ & 2.70 & $0.99 \pm 0.02$ & 52.30 & 1.02 & $1.00 \pm 0.02$ \\
GRB$_{7}$ & 1.99 & $1.38 \pm 0.06$ & 51.72 & 1.44 & $1.33 \pm 0.03$ \\
\enddata
\label{tab:mleC}
\end{deluxetable}

\begin{deluxetable}{cccccccc}
\tablecolumns{8}
\tablecaption{Fit results from \textit{Mod~A} on Real Data}
\tablewidth{0pt}
\tablehead{\colhead{GRB} & \colhead{$z$} & \colhead{$\mean{z_{\rm est}}$} & \colhead{$z{\rm est}$ 95\% HDI} & \colhead{$\mean{N_{\rm rest, est}}$} & \colhead{$N_{\rm rest, est}$ 95\% HDI} & \colhead{$\mean{\gamma_{\rm est}}$} & \colhead{$\gamma_{\rm est}$ 95\% HDI}}
\startdata
080916C & 4.24 &\nodata& \nodata & 51.53 & 51.26 - 51.82 & 1.36 & 1.16 - 1.53 \\
081224 & \nodata & 6.37 & 0.98 - 14.02 & 51.88 & 51.19 - 52.59 & 1.49 & 1.20 - 1.90 \\
090719 & \nodata & 3.80 & 0.35 - 11.66 & 52.00 & 51.23 - 52.89 & 1.51 & 1.25 - 1.85 \\
090809 & \nodata & 5.50 & 0.66 - 13.28 & 51.93 & 51.21 - 52.68 & 1.51 & 1.22 - 1.87 \\
110721A & 3.20 &\nodata& \nodata & 51.58 & 51.25 - 51.91 & 1.43 & 1.22 - 1.63 \\
110920 & \nodata & 8.08 & 2.61 - 14.95 & 51.73 & 51.19 - 52.19 & 1.45 & 1.29 - 1.63 \\
130427A & 0.34 &\nodata& \nodata & 51.27 & 51.22 - 51.31 & 1.45 & 1.37 - 1.54 \\
141028A & 2.33 &\nodata& \nodata & 51.77 & 51.53 - 51.98 & 1.49 & 1.26 - 1.77 \\
\enddata
\label{tab:realA}
\end{deluxetable}

\begin{deluxetable}{cccccccc}
\tablecolumns{8}
\tablecaption{Fit results from \textit{Mod~B} on Real Data}

\tablewidth{0pt}
\tablehead{\colhead{GRB} & \colhead{$z$} & \colhead{$\mean{z_{\rm est}}$} & \colhead{$z{\rm est}$ 95\% HDI} & \colhead{$\mean{N_{\rm rest, est}}$} & \colhead{$N_{\rm rest, est}$ 95\% HDI} & \colhead{$\mean{\gamma_{\rm est}}$} & \colhead{$\gamma_{\rm est}$ 95\% HDI}}
\startdata

080916C & 4.24 &\nodata& \nodata & 51.67 & 51.35 - 52.00 & 1.25 & 1.03 - 1.47 \\
081224A & \nodata & 6.73 & 0.73 - 14.22 & 51.71 & 50.91 - 52.58 & 1.64 & 0.43 - 2.50 \\
090719A & \nodata & 4.84 & 0.46 - 13.21 & 51.87 & 51.16 - 52.67 & 1.73 & 1.14 - 2.39 \\
090809A & \nodata & 6.61 & 0.82 - 14.17 & 51.75 & 51.06 - 52.52 & 1.76 & 1.00 - 2.48 \\
110721A & 3.20 &\nodata& \nodata & 51.61 & 51.09 - 52.17 & 1.41 & 1.01 - 1.75 \\
110920A & \nodata & 7.80 & 2.33 - 14.60 & 51.71 & 51.19 - 52.20 & 1.46 & 1.22 - 1.70 \\
130427A & 0.34 &\nodata& \nodata & 51.26 & 51.21 - 51.31 & 1.46 & 1.36 - 1.55 \\
141028A & 2.33 &\nodata& \nodata & 51.68 & 51.28 - 52.06 & 1.59 & 1.12 - 2.04 \\

\enddata

\label{tab:realB}
\end{deluxetable}
\clearpage

\begin{deluxetable}{cccccccc}
\tablecaption{Fit results from \textit{Mod~C} on Real Data}
\tablecolumns{8}
\tablewidth{0pt}
\tablehead{\colhead{GRB} & \colhead{$z$} & \colhead{$\mean{z_{\rm est}}$} & \colhead{$z{\rm est}$ 95\% HDI} & \colhead{$\mean{N_{\rm rest, est}}$} & \colhead{$N_{\rm rest, est}$ 95\% HDI} & \colhead{$\mean{\gamma_{\rm est}}$} & \colhead{$\gamma_{\rm est}$ 95\% HDI}}
\startdata
080916C & 4.24 &\nodata& \nodata & 51.70 & 51.33 - 52.06 & 1.23 & 0.99 - 1.48 \\
081224A & \nodata & 7.70 & 1.08 - 14.99 & 52.14 & 50.24 - 53.92 & 1.37 & 0.02 - 2.66 \\
090719A & \nodata & 7.61 & 0.95 - 14.96 & 52.35 & 51.09 - 53.47 & 1.56 & 0.90 - 2.17 \\
090809A & \nodata & 7.51 & 0.41 - 14.36 & 52.05 & 50.39 - 53.49 & 1.57 & 0.47 - 2.61 \\
110721A & 3.20 &\nodata& \nodata & 51.62 & 50.65 - 52.61 & 1.41 & 0.74 - 2.03 \\
110920A & \nodata & 7.76 & 1.03 - 14.77 & 51.66 & 50.66 - 52.36 & 1.44 & 1.19 - 1.69 \\
130427A & 0.34 &\nodata& \nodata & 51.26 & 51.21 - 51.31 & 1.46 & 1.37 - 1.56 \\
141028A & 2.33 &\nodata& \nodata & 51.71 & 51.24 - 52.16 & 1.56 & 1.02 - 2.11 \\
\enddata
\label{tab:realC}
\end{deluxetable}

\begin{deluxetable}{cccc}
\tablecaption{{\tt FITEXY Results using all known redshift GRBs as Calibration Sources}}
\tablecolumns{4}
\tablewidth{0pt}
\tablehead{\colhead{GRB} & \colhead{$z$} & \colhead{$z_{\rm est}$} & \colhead{$\gamma_{\rm est}$}}
\startdata
080916C & 4.24 & $2.51 \pm 0.50$ & $1.39 \pm 0.09$ \\
081224A & \nodata & $3.64 \pm 583.90$ & $3.05 \pm 1.38$ \\
090719A & \nodata & $0.88 \pm 0.22$ & $1.80 \pm 0.17$ \\
090809A & \nodata & $1.29 \pm 0.53$ & $1.80 \pm 0.25$ \\
110721A & 3.20 & $1.70 \pm 0.33$ & $1.44 \pm 0.09$ \\
110920A & \nodata & $2.40 \pm 0.34$ & $1.50 \pm 0.07$ \\
130427A & 0.34 & $0.31 \pm 0.02$ & $1.52 \pm 0.05$ \\
141028A & 2.33 & $1.49 \pm 0.68$ & $1.91 \pm 0.24$ \\
\enddata
\label{tab:mleReal}
\end{deluxetable}

\begin{deluxetable}{ccccc}
\centering
\tablecolumns{5}
\tablecaption{Predicted Redshifts via {\tt FITEXY} using individual GRBs as Calibration Sources}
\tablewidth{0pt}
\tablehead{\colhead{Calib. GRB} & \colhead{080916C ($z=4.24$)} & \colhead{110721A  ($z=3.2$)} & \colhead{130427A  ($z=0.34$)} & \colhead{141028A  ($z=2.332$)}}
\startdata
080916C & $4.24 \pm 1.27$ & $2.75 \pm 0.72$ & $0.45 \pm 0.03$ & $2.71 \pm 1.67$ \\
110721A & $5.01 \pm 1.66$ & $3.20 \pm 0.91$ & $0.50 \pm 0.04$ & $3.33 \pm 2.31$ \\
130427A & $2.79 \pm 0.58$ & $1.88 \pm 0.37$ & $0.34 \pm 0.02$ & $1.68 \pm 0.80$ \\
141028A & $3.74 \pm 1.32$ & $2.45 \pm 0.72$ & $0.42 \pm 0.03$ & $2.33 \pm 1.41$ \\
\enddata
\label{tab:mleReal2}
\end{deluxetable}

\clearpage

\appendix

\section{Marginal Posterior Distributions from Simulations}
\label{sec:post}

\begin{figure}
  \centering
  \includegraphics{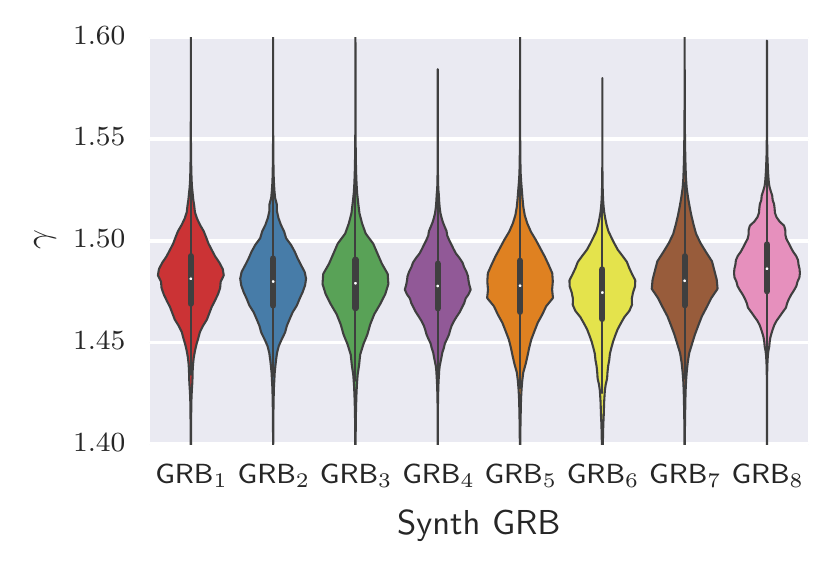}
  \caption{The estimated distributions of $\gamma^i$ for simulations
    of \textit{Mod~A}.}
  \label{fig:simAgamma}
\end{figure}

\begin{figure}
  \centering
  \includegraphics{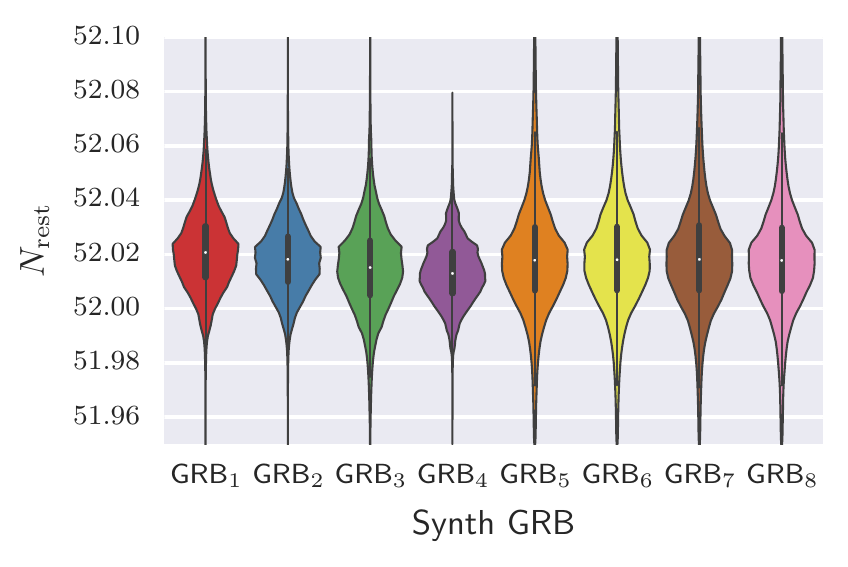}
  \caption{The estimated distributions of $\Nr^i$ for simulations
    of \textit{Mod~A}.}
  \label{fig:simANr}
\end{figure}

\begin{figure}
  \centering
  \includegraphics{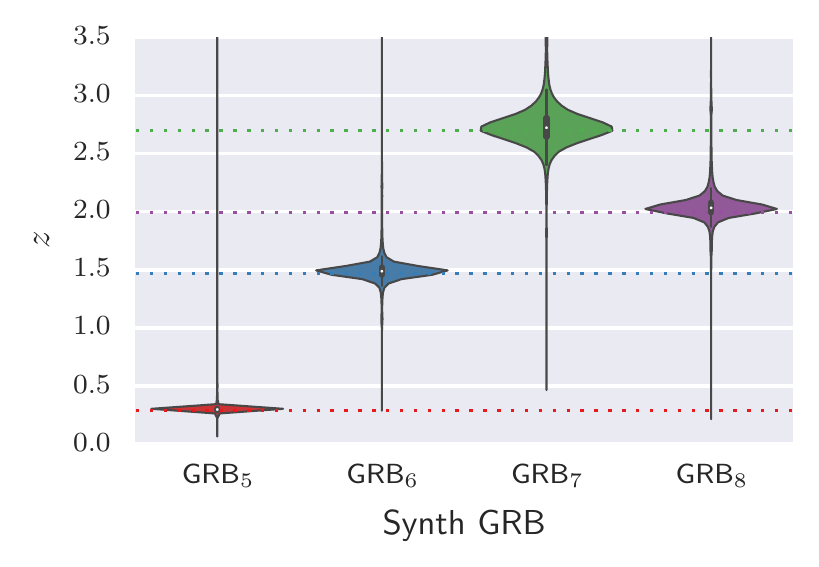}
  \caption{The estimated distributions of unknown redshifts for
    simulations of \textit{Mod~A}. The dashed lines indicate the
    simulated values.}
  \label{fig:simAz}
\end{figure}

\begin{figure}
  \centering
  \includegraphics{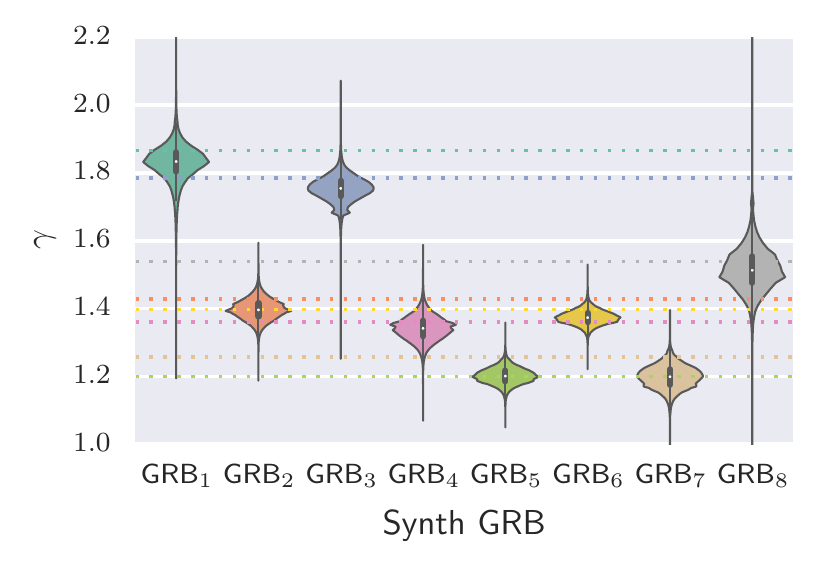}
  \caption{The estimated distributions of $\gamma^i$ for simulations
    of \textit{Mod~B}.}
  \label{fig:simBgamma}
\end{figure}

\begin{figure}
  \centering
  \includegraphics{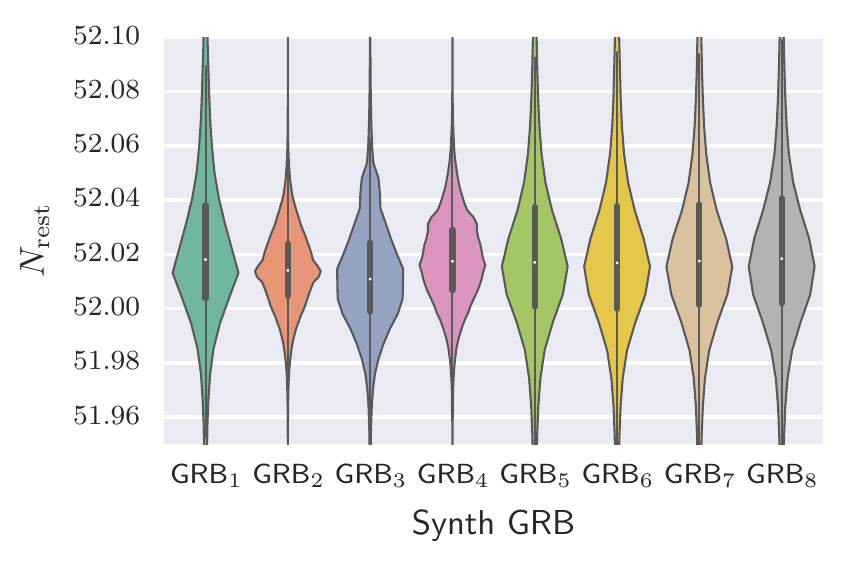}
  \caption{The estimated distributions of $\Nr^i$ for simulations
    of \textit{Mod~B}.}
  \label{fig:simBNr}
\end{figure}

\begin{figure}
  \centering
  \includegraphics{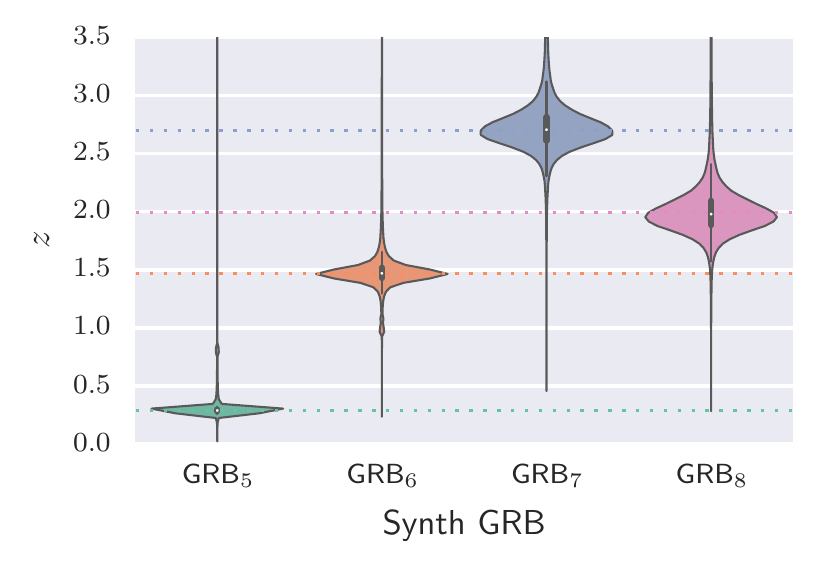}
  \caption{The estimated distributions of unknown redshifts for
    simulations of \textit{Mod~B}. The dashed lines indicate the
    simulated values.}
  \label{fig:simBz}
\end{figure}

\begin{figure}
  \centering
  \includegraphics{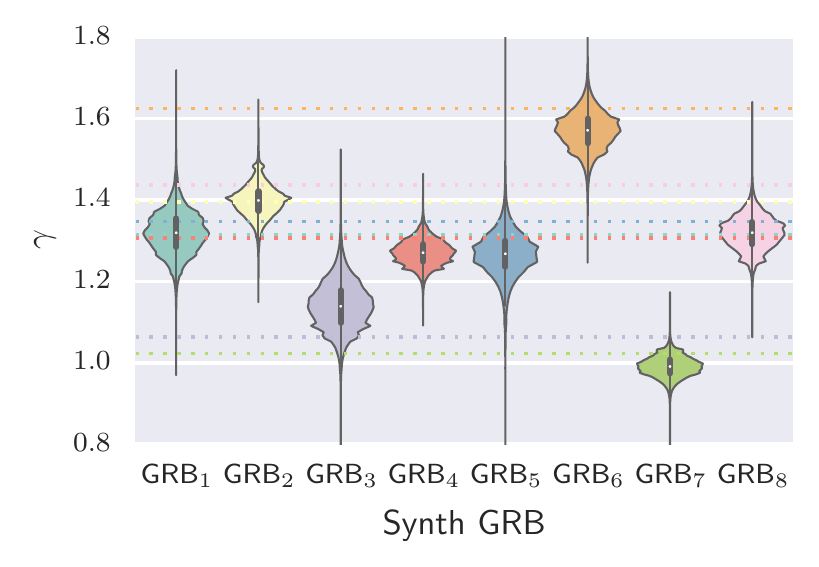}
  \caption{The estimated distributions of $\gamma^i$ for simulations
    of \textit{Mod~C}.}
  \label{fig:simCgamma}
\end{figure}

\begin{figure}
  \centering
  \includegraphics{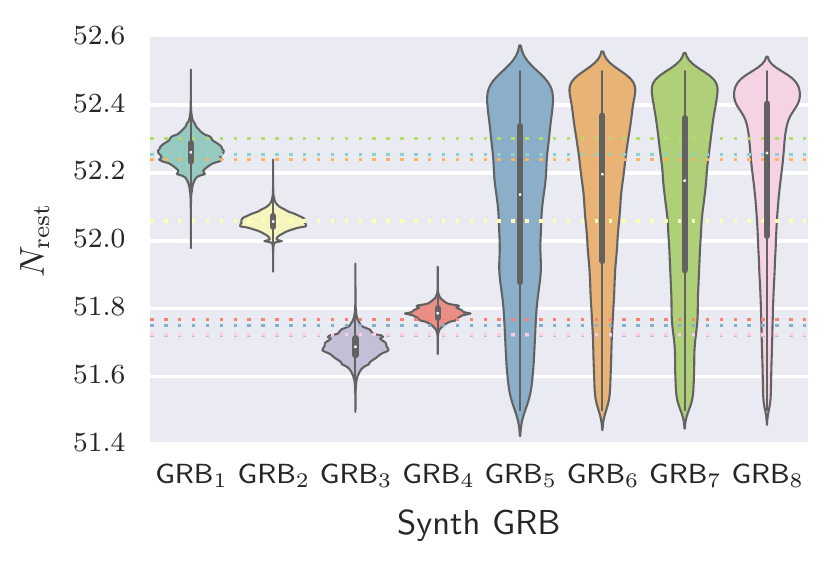}
  \caption{The estimated distributions of $\Nr^i$ for simulations
    of \textit{Mod~C}.}
  \label{fig:simCNr}
\end{figure}

For completeness, I display the marginal posterior distributions of
the simulations. Figures \ref{fig:simAgamma}-\ref{fig:simAz} display
the distributions from simulations of \textit{Mod~A}. Figures
\ref{fig:simBgamma}-\ref{fig:simBz} display the distributions of
simulations from \textit{Mod~B}. Finally, Figures
\ref{fig:simCgamma}-\ref{fig:simCNr} display the distributions of
simulations from \textit{Mod~C}.

\section{$\FE$ Error Propagation}
\label{sec:prop}

In order to calculate the errors on $\FE$ for a single component from
a multi-component fit, one must recognize that in {\tt RMFIT}, it is
the total model that is fitted and therefore care must be taken to
compute the individual component errors. Suppose we have a
multi-component photon model fitted to data:

\begin{equation}
\mathcal{M}_{\rm T}\left(\varepsilon ; \vec{ \theta_{\rm T}} \right) = \mathcal{M}_{A}\left(\varepsilon ; \vec{\theta_{\rm A}} \right) +  \mathcal{M}_{B}\left(\varepsilon ; \vec{\theta_{\rm B}} \right) \; {\rm pht \; s}^{-1} {\rm cm}^{-2} {\rm keV}^{-1}
\end{equation}

\noindent where $\mathcal{M}_{\rm T}$, $\mathcal{M}_{\rm A}$, and
$\mathcal{M}_{\rm B}$ are the total photon model, and photon models A and B
respectively with parameter vectors $ \vec{\theta_{i}}$. Since we
measure $\mathcal{M}_{\rm T}$, then computing $F_{\rm E, A}$ is
equivalent to

\begin{equation}
  \label{eq:4}
  F_{\rm E, A} = \int {\rm d}\varepsilon \; \left[\varepsilon \mathcal{M}_{\rm T}\left(\varepsilon ; \vec{ \theta_{\rm T}} \right) - \mathcal{M}_{\rm B}\left(\varepsilon ; \vec{ \theta_{\rm B}} \right)\right] {\rm keV \; s}^{-1} {\rm .}  
\end{equation}

\noindent Hence, the errors on the single A component are correlated to
the total model and second B component. Let ${\sigma_{{\rm T},ij}}^2$ be
the covariance matrix from the spectral fit, $A_i = \frac{\partial
  F_{{\rm E, A}}}{\partial
  \theta_i}\vert_{\mean{\theta_i}}$. Proceeding with the typical
Taylor expansion error propagation scheme, the variance of $ F_{\rm E,
  A}$ is
\begin{equation}
  \label{eq:6}
  \sigma_{ F_{\rm E, A}}^2 =\sum_i \sum_j A_i \sigma_{{\rm T},ij}^2 A_j {\rm .}
\end{equation}

If one calculates the error on a component's $\FE$ by considering only
the errors on its parameters, then the errors $\FE$ become much larger
than the method demonstrated here which can lead to unrealistic
flexibility when fitting GCs by either the Bayesian model or {\tt
  FITEXY}. To demonstrate the difference in the techniques we compare
the error of the non-thermal component $\FE$ for the fourth time
interval of GRB080916C from both methods. The $\FE = 3.14\cdot
10^{-6}$ erg s$^{-1}$ and $\sigma_{\FE}=4.53 \cdot 10^{-7}$ erg
s$^{-1}$ from the method demonstrated here and $\sigma_{\FE}=1.18
\cdot 10^{-6}$ erg s$^{-1}$ from the method that does not take into
account the full parameter correlations. The order of magnitude
difference may explain the differences between Figure
\ref{fig:realdata} herein and Figure 15 of \citet{Guiriec:2015}.

\section{Stan Model Code}

The following code demonstrates the Stan implementation of \textit{Mod
  A}. The code takes the known $\Epo$ and $\FE$ with their associated
errors ({\tt Ep\_obs}, {\tt FE\_obs}, {\tt Ep\_obs}, {\tt FE\_sig}) as
well as the pre-calculated rest-frame quantities ({\tt Nz}) and known
redshifts ({\tt z\_known}). A warmup/sample split of 1/2 is taken with
30K iterations which insured convergence.

\begin{lstlisting}


data{
     int<lower=0> N;
     int<lower=0> Nunknown;
     int<lower=0> Ngroups;
     vector[N] z_known;
     vector[N] Nz;
     int group[N];
     int unknown_group[N];
     int known[N];     
     real maxSlope;   
     vector[N] FE_obs; 
     vector[N] Ep_obs; 
     vector[N] Ep_sig; 
     vector[N] FE_sig; 
    }


parameters {

     vector<lower=-2,upper=5>[N] Ep_true;
     vector[N] FE_true;
     vector<lower=50>[Ngroups] Nrest;
     vector<lower=0>[Ngroups] gamma;  
     vector<lower=0.0,upper=15>[Nunknown] z;
     vector<lower=0>[Ngroups] int_scatter_sq; 
     real<lower=0> gamma_mu_meta;
     real Nrest_mu_meta;
     real<lower=0> gamma_sig_meta;
     real<lower=0> Nrest_sig_meta;
  
 }



transformed parameters {
    
    vector<lower=0>[Ngroups] int_scatter;
    
    for(i in 1:Ngroups){ 
       int_scatter[i] <- sqrt(int_scatter_sq[i]);
    }   
 }


model {
  
       gamma_sig_meta ~ cauchy(0.,2.5);
       Nrest_sig_meta ~ cauchy(0.,2.5);
       gamma_mu_meta  ~ normal(0,maxSlope);
       Nrest_mu_meta  ~ normal(52,5);
       gamma  ~ normal(gamma_mu_meta, gamma_sig_meta);
       Nrest  ~ normal(Nrest_mu_meta, Nrest_sig_meta);
       int_scatter_sq ~ cauchy(0,2.5)
       z ~ uniform( 0, 15);
       Ep_true ~ uniform(-2,5);
       Ep_obs ~ normal(Ep_true,Ep_sig); 
    
       for(i in 1:N){
    
          if (known[i] == 0){             
            FE_true[i] ~ normal( Nrest[group[i]] -
            ( 1.099 +
            2*log10( DL( z[unknown_group[i]]))) +
            gamma[group[i]]* 
            (log10(1+z[unknown_group[i]]) +
             Ep_true[i]-2), int_scatter[group[i]]);
           
             }
             
            else{
              FE_true[i] ~ normal(Nrest[group[i]] -
              Nz[i] + gamma[group[i]] * 
              (log10(1+z_known[i]) +
              Ep_true[i]-2), int_scatter[group[i]]);
           
         }
        }
        FE_obs ~ normal(FE_true,FE_sig);
} 



\end{lstlisting}

\clearpage

\label{lastpage}

\end{document}